%% file: main.tex
\documentclass[journal]{IEEEtran}

\input{additional_packages}

\title{Edge-Only Universal Adversarial Attacks in Distributed Learning}

\author{Giulio Rossolini, Tommaso Baldi, Alessandro Biondi and Giorgio Buttazzo \\
Department of Excellence in Robotics \& AI, \\ Scuola Superiore Sant'Anna, Pisa, Italy\\
{\tt\small name.surname@santannapisa.it}
}

\begin{document}
\maketitle

\begin{abstract}
\add{Distributed learning frameworks, which partition neural network models across multiple computing nodes, enhance efficiency in collaborative edge-cloud systems, but may also introduce new vulnerabilities to evasion attacks, often in the form of adversarial perturbations. In this work, we present a new threat model that explores the feasibility of generating universal adversarial perturbations (UAPs) when the attacker has access only to the edge portion of the model, consisting of its initial network layers. Unlike traditional attacks that require full model knowledge, our approach shows that adversaries can induce effective mispredictions in the unknown cloud component by manipulating key feature representations at the edge.
Following the proposed threat model, we introduce both edge-only untargeted and targeted formulations of UAPs designed to control intermediate features before the split point. Our results on ImageNet demonstrate strong attack transferability to the unknown cloud part, and we compare the proposed method with classical white-box and black-box techniques, highlighting its effectiveness.}
Additionally, we analyze the capability of an attacker to achieve targeted adversarial effects with edge-only knowledge, revealing intriguing behaviors across multiple networks.
By introducing the first adversarial attacks with edge-only knowledge in split inference, this work underscores the importance of addressing partial model access in adversarial robustness, encouraging further research in this area.

\end{abstract}

\input{introduction}
\input{rel}

\input{methodology}
\input{exp}

\input{conclusion}

{\small
\bibliographystyle{IEEEtran}
\bibliography{main}
}

\appendix
\input{appendix}
\end{document}

%% file: additional_packages.tex
\usepackage{tikz}

\usepackage{amsthm, amssymb, amsmath}
\usepackage{hyperref}
\usepackage{multirow}%
\usepackage{pifont}
\usepackage{listings}
\usepackage{xcolor}
\usepackage{lipsum}
\usepackage{booktabs}%
\usepackage[export]{adjustbox}
\usepackage{wrapfig} 
\usepackage{graphicx}
\usepackage{subcaption}
\usepackage{pdflscape}
\usepackage{xspace}
\usepackage{breqn}
\usepackage{algorithm}
\usepackage{algpseudocode}
\usepackage{colortbl} 
\usepackage{rotating}
\usepackage{orcidlink}
\usepackage{cleveref}

\definecolor{lightred}{RGB}{255, 220, 220}
\definecolor{lightgrey}{RGB}{240,240,240}

\newcommand {\add}[1]{{\color{black}{#1}}}

\theoremstyle{definition}%

%% file: introduction.tex
\section{Introduction}
\label{sec:intro}
In the last decade, adversarial attacks have posed significant challenges in deploying secure and robust deep learning models \cite{cina2023wild, Carlini017, Szegedy14}. These attacks involve carefully crafted perturbations to input data that can effectively mislead model predictions, originating serious concerns about the safety and security of machine learning in various application domains, such as autonomous driving \cite{rossolini_tnnls_2023, zhang2022adversarial} and healthcare \cite{KAVIANI2022116815}.

From a practical perspective, the severity of these attacks depends on the form of perturbation used by the attacker and on the threat model under investigation. 
Regarding the former, different types of attacks exist, including classic input-specific perturbations \cite{Szegedy14, pgd_attack, Carlini017}, universal adversarial perturbations (UAPs) \cite{zhangsurveyijcai, MoosaviUAPCVPR, Zhang_data-free}, and physical attacks \cite{adversarialtshirt_Xu_2020, brown_adversarial_2018, rossolini_iccps24}. Additionally, an attack can either be crafted to follow a targeted objective, where the model is forced to predict a specific class, or an untargeted objective, which instead aims at a misclassification irrespective of the predicted class.


In terms of the threat model, adversarial attacks and the robustness of deep neural networks have been mostly studied in centralized environments, where the model is either fully accessible to the attacker (white-box) \cite{Szegedy14} or entirely inaccessible in a black-box setting \cite{Bhambri2019ASO}. However, with the rise of distributed learning systems \cite{split_survey,Neurosurgeon}, where inference is split across multiple nodes, such as edge and cloud ones, new threat scenarios are emerging. As in this context attackers may gain access to a portion of the model only, a deeper investigation of the threats to which distributed learning is exposed becomes fundamental \cite{security_imp_edge_comp, guo2021robust_collaborative}.

Previous work on robustness of distributed learning has primarily focused on threats related to the training phase, where models can be susceptible to attacks mounted at training time \cite{Yu_Zeng_Zhao_Pang_Wang_2024, Cao2019UnderstandingDP} and the attacker controls (a part of) the cloud node \cite{Liu2022SimilarityBasedLI,paquini_split_tiger,erdougan2022unsplit}. In contrast, motivated by the many hardware and system-level attacks that allow compromising edge nodes (even by means of physical access)\cite{ecs_security, ecs_security2}, this work addresses a different but yet practically-relevant threat model: \emph{the attacker can control the edge part only at the deployment (i.e., post-training) stage}. 
Designing attacks in this context presents unique challenges as the attacker lacks access to the cloud part of the model, the final outputs, and even full knowledge of the class distribution, complicating the generation of effective adversarial attacks.

\add{
To address these challenges, this work formalizes a novel edge-only adversarial threat model and proposes methods for generating UAPs in both targeted and untargeted settings. For the targeted formulation, the main idea is to exploit a small auxiliary attack set to learn class-specific feature patterns from the accessible edge layers, by contrasting a reference target class against all other classes. For the untargeted variant, the objective is instead to induce deviations from clean feature representations.
Despite the lack of access to the remaining model components, we show that edge features are sufficient to guide the optimization of universal perturbations, steering attacked inputs to cause misclassification when propagated through the unknown cloud-side model.
}

\add{
We conducted extensive experiments on ImageNet \cite{NIPS2012_c399862d}, varying both the number of layers deployed on the edge and the model architectures. The results show that the proposed attack paradigm induces high misclassification rates, even when only a few layers are accessible at the edge, achieving performance often comparable to white-box UAP methods \cite{data_free_zhang_ICCV2021, MoosaviUAPCVPR} in the untargeted setting. Comparable effectiveness is also observed for the targeted formulation, depending on the chosen target class and the depth of the accessible edge portion.
}
Additionally, we provide an in-depth analysis of how targeted optimization propagates through the unknown cloud portion of the model, revealing that even when the attack is applied at the initial layers only, it shifts features towards specific unintended classes on the cloud side, effectively mimicking a targeted attack but towards a different class.
\add{
In summary, this work makes the following contributions:
\begin{itemize}
    \item It introduces an edge-only threat model for distributed inference, where the cloud part of the model is unknown;
    \item It presents formulations of both targeted and untargeted edge-only UAPs under the proposed threat model;
    \item It reports extensive experimental results comparing the effectiveness of edge-only UAPs with traditional full-knowledge and black-box attacks, highlighting the practical relevance of addressing the proposed threat model.
\end{itemize}
}
\add{
The remainder of the paper is organized as follows: Sec.~\ref{s:rel} discusses related work; Sec.~\ref{s:background} introduces background concepts; Sec.~\ref{s:threat_model} presents the threat model; Sec.~\ref{s:attack} details the proposed methodologies for crafting edge-only universal adversarial attacks, first in the targeted and then in the untargeted setting; Sec.~\ref{s:exp} presents the experimental results; and finally, Sec.~\ref{s:conc} presents the conclusions and directions for future work\footnote{\add{Code available at \url{https://github.com/retis-ai/EdgeUAP}}}.

}

%% file: rel.tex
\section{Related Work}
\label{s:rel}
In the following, we draw attention to our work by discussing related studies from two perspectives: universal adversarial attacks and the security threats posed within distributed learning scenarios.

\subsection{Universal Adversarial Attacks} 
In recent years, adversarial perturbations have gained significant attention as a way to reveal critical weaknesses in the robustness of DNNs \cite{Szegedy14, Carlini017, pgd_attack}. Among these, universal adversarial attacks are particularly important from a security standpoint, as they maintain their effectiveness across different inputs \cite{UAP_trainigg, metzne_UAP_SS, MoosaviUAPCVPR, zhangsurveyijcai}.
\add{This enables attackers to generate the perturbation offline and apply it at runtime, making such attacks significantly more practical in real-world scenarios.} 
Moosavi-Dezfooli et al.\cite{MoosaviUAPCVPR} were the first to demonstrate the existence of UAPs by introducing an adversarial optimization method that extends the DeepFool attack\cite{Moosavi16_deepfool} to extract features that generalize across multiple images. Since then, various strategies have been developed to enhance UAPs' effectiveness \cite{Mopuri2017FastFF, AnalysisDezfooli17}. 
In these studies, also targeted formulations of UAPs were explored, which are practically interesting as they avoid the need for labeled samples during optimization (i.e., the perturbation is optimized to push any input towards a specific class) \cite{benz2020double, MoosaviUAPCVPR}.
In addition, interesting studies have been conducted to understand the nature of UAPs, suggesting that they often relate to task-specific features rather than the training data itself \cite{data_free_zhang_ICCV2021, zhang2021universal, MopuriGANUAP, naseer2019cross}, and, when applied, these features become predominant across layers\cite{data_free_zhang_ICCV2021}. 

Despite the extensive literature on UAPs, these techniques typically assume full, white-box access to the model, while scenarios involving partial knowledge of the weights and architecture remained largely unexplored. In this context, we address the challenge of determining whether addressing specific shallow features that correlate with target classes could help craft UAPs capable of inducing misclassification across the remaning part of the model.

\subsection{Threat Models in Distributed Learning}
The development of distributed and collaborative learning paradigms has enabled the training and inference of AI models across multiple nodes, improving computational efficiency in edge computing environments and IoT applications \cite{split_survey, Neurosurgeon}. However, these distributed paradigms also introduces novel security vulnerabilities and broadens the threat landscape \cite{security_imp_edge_comp}. Prior research has explored various threats in distributed settings, including backdoor attacks in federated learning \cite{lyu2022privacy, Xie_distributed_backdoor_attacks_2020} and privacy issues in split learning through manipulation of cloud-side components \cite{paquini_split_tiger, erdougan2022unsplit}.

From the perspective of evasion attacks, however, previous work has not yet explored scenarios beyond conventional adversarial settings, where attackers rely on access to model outputs, either through full model access in white-box attacks or in black-box attacks \cite{Papernot2017Practical, Yin_tpami_2024, Xu_tifs, Ma_tifs_2023}.
To fill this gap, we investigate a previously underexplored but realistic threat scenario in distributed model partitioning: an attacker gaining control over the edge component during inference. As discussed in more detail in Sec.\ref{s:background}, this threat model is particularly relevant due to the practical limitations of edge devices, which often operate with constrained computational resources and weak security measures. These characteristics make them vulnerable to model cloning, parameter extraction, and gradient analysis at the partitioning point \cite{Dong2022SplitNetsDN, security_imp_edge_comp, Sharma_stealing_2023}.

To address this gap, we study how an attacker can exploit shallow representations using UAPs, leveraging only the feature distribution from the edge component without access to the full model or decision boundary. This approach enables new insights into the vulnerability of distributed inference under realistic deployment constraints.

%% file: methodology.tex
\section{Background}
This section recaps key terminologies related to split inference in distributed learning and universal perturbations.
\label{s:background}

\subsection{Split Inference.}
In split inference \cite{split_survey, Neurosurgeon}, a DNN \( f \) is divided into distinct portions. For simplicity, we consider DNNs split into an \textit{edge part} \( f_{\text{edge}} \) and a \textit{cloud part} \( f_{\text{cloud}} \), such that $f(x) = f_{\text{cloud}}(f_{\text{edge}}(x))$. 
In this work, $f$ is an image classifier, as commonly addressed in the UAP literature. Nevertheless, the whole approach and terminology can be seamlessly extended to other tasks. The edge part is composed by the first $e$ layers of the model, i.e.,  $L_{\text{edge}}= \{l_1, l_2, ..., l_e\}$, and processes an input \( x \) by returning intermediate features \( h^{l_e} = f_{\text{edge}}(x) \). These features \( h^{l_e} \) are then transmitted to the cloud part, which completes the remaining inference to generate the final model output \( f(x) =  f_{\text{cloud}}(h^{l_e})\).
When needed, we also denote by $h^{l_i} = f^{l_i}(x)$ the output of the model up to layer $l_i$.

\subsection{Universal Adversarial Perturbations.}
Formally, the objective is to find a perturbation \( \delta \) that induces mispredictions on any input \( x \) of a distribution \( X \). This problem can be formulated as follows:
\begin{equation}\label{eqn:UAP-general}
\max_{\delta} \; \mathbb{E}_{x \sim X} \left[ \mathcal{L}(f(x + \delta), y) \right] \quad \text{s.t.} \quad \|\delta\|_p \leq \epsilon,
\end{equation}
where $y$ is the ground-truth output related to $x$, \( \mathcal{L} \) is a loss function (e.g., cross-entropy loss), and \( \epsilon \) is the magnitude of \( \delta \) under a \( p \)-norm constraint.
In this work, we focus on the $l_\infty$ norm, commonly adopted in the related literature \cite{zhang2021universal}.

Particularly relevant to this work is the targeted variant of UAPs, where the goal is to induce predictions towards a specific target class \( y_t \) selected by the attackers, i.e., minimizing a loss function with respect to \( y_t \): 
\begin{equation} \label{eqn:uap-general}
\min_{\delta} \; \mathbb{E}_{x \sim X} \left[ \mathcal{L}(f(x + \delta), y_t) \right] \quad \text{s.t.} \quad \|\delta\|_p \leq \epsilon,
\end{equation}

Let \( D_{\text{opt}} \) be a dataset collected by the attacker that does not include inputs of the target class $y_t$ only, and without any particular restriction on its distribution. In practice, the perturbation \( \delta \) can be iteratively computed by means of gradient:
\begin{equation}
\delta_{t+1} = \delta_t - \alpha \cdot \text{sign}\left( \sum_{x \in B} \nabla_{\delta} \mathcal{L}(f(x + \delta_t), y_t) \right), 
\end{equation}
where \( B \) represents a batch of samples from \( D_{\text{opt}} \), \( \alpha \) is the step size, and $\textit{sign}$ the $l_\infty$ projection. After each update  a \textit{clip} in $[-\epsilon,+\epsilon]$ is applied, to ensure the $\epsilon$ constraint.

\section{Edge-Only Attack Threat Model}
\label{s:threat_model}

We study a secure setting in which the model output is not returned to the edge part \cite{paquini_split_tiger}.
It assumes that: \textit{(i)} the attacker can collect a small set of samples that differ from the training set; \textit{(ii)} the attacker has white-box access to the edge portion of the model only, i.e., \( f_{\text{edge}} \), allowing to study its activations and perform gradient backpropagation;  \textit{(iii)} the complete set of output classes \( \{1, \dots, N_c\} \) is unknown to the attackers, while they can focus on a specific class (the target class) that is assumed to be part of this set (but not strictly necessary).
In this setting, the attacker’s goal is to craft an adversarial perturbation that causes the model (on the cloud side) to produce an incorrect prediction.

The practical relevance of the above assumptions is motivated as follows: \textit{(i)-(ii)} the attacker can gain access to the edge node, even by means of physical access, hence being able to leak both the edge portion of the model and input samples; \textit{(iii)} the attacker can classify a small set of samples independently of the model $f$ under attack.

\add{
\subsection{Relevance of the Threat Model}
The proposed threat model captures a setting in which the attacker has no access to model outputs and no knowledge of the full label space, but can exploit internal representations from an initial portion of the network (the edge part). This scenario is increasingly relevant due to the growing deployment of split and collaborative neural networks in sensor nodes and IoT systems \cite{Teerapittayanon2017DistributedDNN, Neurosurgeon} (e.g.,  wearable health-monitoring devices, smart retail sensors), where resource-constrained edge devices perform early encoding for feature extraction, compression, and partial inference, while the remaining computation is offloaded to remote servers.
For instance, representative examples are smart video surveillance and traffic monitoring, where edges perform on-device feature extraction and transmit intermediate embeddings to a cloud service for final recognition or analytics. In such systems, the edge device is physically exposed and can be accessed, probed, or tampered with by an adversary, while the cloud model and final predictions remain protected behind secure infrastructure. 
In these practical deployments, compromising the edge-side model or its activations is significantly more feasible than breaching the cloud backend \cite{Sharma_stealing_2023,ecs_security, ecs_security2}, reflecting a realistic and underexplored attack surface that emerges in modern distributed inference pipelines.
}

\begin{figure}[t]
\begin{center}
\begin{subfigure}{\columnwidth}
\includegraphics[width=\textwidth]{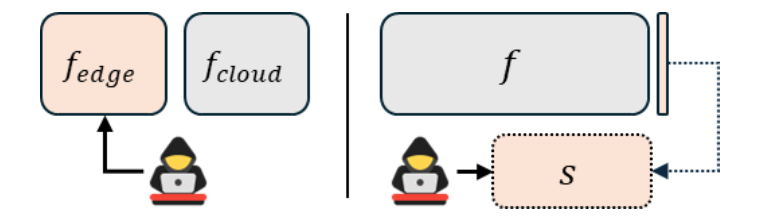}
\end{subfigure}
\vspace{-1em}
\caption{
 \small{\add{Comparison between threat models. Left: the proposed edge-based attack, where the adversary has access to the edge portion of the split model. Right: a standard black-box setting, where the adversary relies on a surrogate model $s$ to craft transferable attacks.}}}
\label{f:comparisons_t_models}
\end{center}
\end{figure}

\add{
\subsection{Clarifications with respect to black-box attacks.}
The presented threat model does not fall under the standard \textit{weak black-box} setting. As illustrated in Fig.\ref{f:comparisons_t_models}, classical black-box attacks typically assume that the adversary has access to the model’s outputs (e.g., in the form of predicted labels or full confidence scores), while the internal architecture and parameters remain hidden. In such settings, it is common to assume that the attacker can train or acquire a surrogate model ($s$ in Fig.\ref{f:comparisons_t_models}), which is then used to craft adversarial examples that are expected to transfer to the target \cite{AnalysisDezfooli17, nakka_indirect_2020, brau2025transferbench}.

These assumptions, although widely adopted to evaluate transferability across architectures, are often optimistic in practice: in private and proprietary deployments, the attacker may lack access to sufficiently representative data or architectural priors to effectively train highly transferable surrogates. 

In contrast, our work focuses on a fundamentally different scenario in which the input is processed locally at the edge and no prediction feedback is returned to the attacker. This setup cannot be reduced to a standard black-box model, as the attacker has access to internal parameters (up to the split point) but no access to output logits or labels.
}

\section{Edge-Only Universal Attack}
\label{s:attack}

\begin{figure*}[t]
\begin{center}
\includegraphics[width=\textwidth]{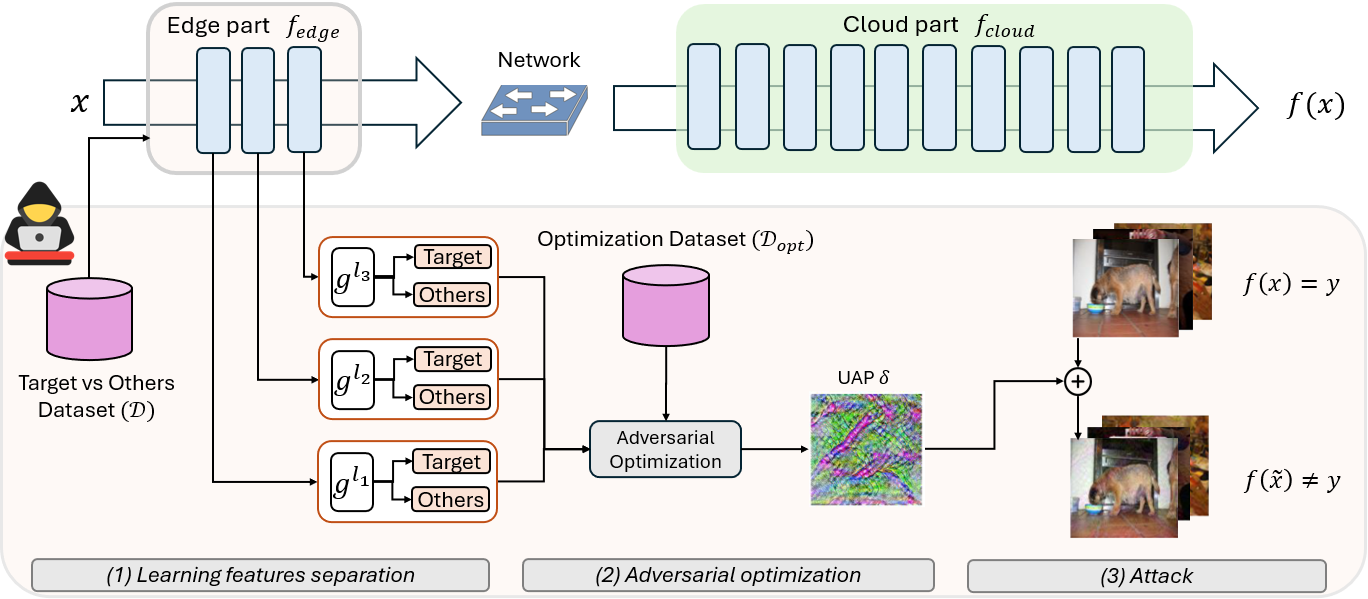}
\caption{
\small{\add{Scheme of the proposed targeted edge-only universal attack (Secs. \ref{ss:learning_class} and \ref{ss:opt_uap}). The attacker has full control only over the edge part, from which can exploit the output of its layers to train binary classifiers $g^l$ that learn feature separations characterizing reference target samples. This is then exploited to run an adversarial optimization that crafts a universal attack transferable to the cloud part of the model.}}}
\label{f:scheme_attack}
\end{center}
\end{figure*}

\add{In this section, we detail the proposed edge-only universal attacks. We first formulates a targeted attack, following the pipeline illustrated in Fig.~\ref{f:scheme_attack}. The method consists of two main stages: (i) learning feature-level separations between target and non-target samples in the edge layers (Sec.~\ref{ss:learning_class}); and (ii) crafting a universal perturbation by exploiting these learned separations (Sec.~\ref{ss:opt_uap}). 
We then introduce, in Sec.~\ref {ss:untarget_attack}, an untargeted attack that causes misclassification on the cloud side without controlling the predicted class.
}

\subsection{Learning Class-Feature Separations for Targeted Attacks}
\label{ss:learning_class}
In this stage, we aim to identify which intermediate features best represent the target class distribution. To this end, we assume that the attacker has access to a small set of samples linked to the target class, denoted by \( \mathcal{D}_t \) (e.g., 50 samples), and a complementary set of samples from other classes, denoted by \( \mathcal{D}_o\), with no particular restrictions on its distribution. We refer to the combined dataset as \( \mathcal{D} = \{\mathcal{D}_t \cup \mathcal{D}_o\} \). Note that the attacker does not require detailed information about non-target samples: only binary labels indicating whether a sample belongs to the target class or not are required.

Given the access to the edge part of the model, \( f_{\text{edge}} \), the attacker can extract features \( h^l \) from any intermediate layer \( l \) within \( f_{\text{edge}} \). For each such layer \( l \in L_{\textit{edge}} \), the attacker trains a binary classifier \( g^l : \mathbb{R}^{\mathcal{H}^l} \rightarrow [0,1] \), where \( \mathcal{H}^l \) is the dimension of the feature space at layer \( l \), to discriminate the most representative features \( h^l \) between the target and non-target classes. Formally, $g^l$ is trained as follows: 
\begin{equation}
\min_{\theta_{g^l}} \; \mathbb{E}_{x \sim \mathcal{D}} \left[ \mathcal{L}_b\left( g^l(f^l(x)), \bar{y}_t \right) \right],
\end{equation}
\noindent where \( \mathcal{L}_b \) denotes the binary cross-entropy loss, \( \theta_{g^l} \) are the weights of \( g^l \), and \( \bar{y}_t \) is an auxiliary label set to 1 for target samples and to 0 for non-target samples.
By this binary classifier the attacker learns which features most strongly represent the target class and use them to guide the optimization.

It is also important to note that the \emph{depth} of the edge part of the model, defined as the number of layers $e$ deployed at the edge, plays a significant role in the generalization capability beyond the samples in $\mathcal{D}$. In particular, while knowledge of deeper layers improves the transferability of attacks to the cloud part, it also carries a higher risk of overfitting when considering few samples, reducing the ability to generalize.
To address this problem, we found that learning feature separation from multiple edge layers and combining them in an optimization process is beneficial, as also shown in the ablation studies reported in Sec. \ref{s:exp}.

\subsection{\add{Targeted Adversarial Attack}}
\label{ss:opt_uap}
In the adversarial optimization stage, we aim to craft a universal perturbation that pushes an arbitrary sample $x$ to be classified in the target class $y_t$ by the binary classifiers $\{g^{l_1}, \dots, g^{l_e}\}$ introduced above, so that $g^l(f^l(x + \delta))=1$ for $l\in L_{\textit{edge}}$. This problem can be framed as a targeted universal optimization, where the objective is to craft malicious input patterns so that the perturbed features match those of the target class, thereby influencing the processing performed by the cloud part of the model.
Extending the optimization problem of Eq.~\eqref{eqn:uap-general}, a direct approach to address this problem consists of adopting gradient ascent to each classifier $g^l$ as follows:
\begin{equation}
\delta_{t+1} = \delta_t + \alpha \cdot \text{sign}\left(  \sum_{l\in L_{\text{edge}}} \sum_{x \in B} \nabla_{\delta} g^l(f^l(x+\delta_t)) \right) ,
\end{equation}
\noindent where $\|\delta\|_\infty \leq \epsilon$. Note that we omit the loss function in this formula, as the gradient itself points directly towards the target label \(\bar{y}_t=1\).
However, we observed that the impact of the gradients is not properly balanced, as they are computed at different stages of the model. To address this issue, we balance the influence of multiple binary classifiers \( g^l \) in the adversarial optimization by normalizing each respective gradient and combining them in the function \( \Phi \):
\begin{equation}\label{eqn:phi}
\Phi(x, \delta) = \sum_{l\in L_{\text{edge}}} \frac{\nabla_{\delta} g^l(f^l(x+\delta_t))}{\|\nabla_{\delta} g^l(f^l(x+\delta_t))\|_2},
\end{equation}
where each gradient \( \nabla_{\delta} g^l(\cdot) \) of classifier \( g^l \) is normalized by its \( \ell_2 \)-norm to ensure balanced contributions across layers. Both the usage of multiple gradients and the normalization allow improving the transferability of the attack to the cloud part, as shown in studies reported in Sec. \ref{ss:ablation}. 

The optimization is performed by iteratively updating the perturbation \( \delta \) in the direction of \( \Phi \), which pushes the intermediate features to resemble those of the target class across all edge layers, that is
\begin{equation}
\delta_{t+1} = \delta_t + \alpha \cdot \text{sign}\left(  \sum_{x \in B} \Phi(x, \delta_t) \right) ,
\label{eq:opt_attack}
\end{equation}

\noindent where $\|\delta\|_\infty \leq \epsilon$ keeps the perturbation under the $\epsilon$-magnitude. 

The resulting perturbed samples can eventually be used to induce mispredictions, keeping their representation in $[0,1]$, i.e., $\tilde{x} = \textit{clip}(x+\delta , 0, 1)$,  assuming images are normalized between 0 and 1 for convenience.


\add{
As observed in the experimental part (Sec.\ref{s:exp}), the feasibility of achieving high targeted misclassification in edge-only settings depends on the degree of separation of class feature representations available at the edge depth. This characteristic depends on the model architecture, the selected class, and the depth of the accessible edge layers, which may exhibit limited class separability.
Nevertheless, we found that such perturbations can effectively influence the final predictions, leading to misclassification with respect to the correct label, i.e., \( f(x + \delta) \neq y \), even when the perturbation does not consistently drive the prediction toward a specific target class. This observation motivates the definition of the next attack variant, which is specifically designed for untargeted objectives.
}

\subsection{\add{Untargeted Adversarial Attack}}
\label{ss:untarget_attack}
\add{
We also introduce an untargeted adversarial formulation as a direct extension of the CosUAP method by Zhang et al.~\cite{data_free_zhang_ICCV2021}. While CosUAP optimizes a universal perturbation by minimizing, in expectation, the cosine similarity between the output logits of clean and perturbed samples, we adapt this principle to the intermediate features extracted by the edge model and backpropagate gradients only through the white-box accessible layers. Formally, this can be expressed as:

\begin{equation} \label{eqn:cosuap_edge}
\min_{\|\delta\|_p \leq \epsilon ,} \; \mathbb{E}_{x \sim X} \left[ 
\mathcal{L}_{\text{cos}}(f_{\text{edge}}(x + \delta), f_{\text{edge}}(x))
\right] ,
\end{equation}

\noindent where 
$\mathcal{L}_{\text{cos}}$ denotes the cosine similarity. In practice, $\delta$ is updated iteratively, as in the previous targeted case (Eq.~\ref{eq:opt_attack}), but without the need to train the auxiliary classifier \( g \) here, since no class-specific feature learning is required. Following the approach in Eq.\ref{eqn:phi}, the update direction is defined here as:

\begin{equation} \label{eqn:phi_unt}
\Phi_{\text{unt}}(x, \delta) = 
- \sum_{l \in L_{\text{edge}}} 
\frac{\nabla_{\delta}\, \mathcal{L}_{\text{cos}}\!\left(f^{l}(x+\delta_t), f^{l}(x)\right)}
{\left\|\nabla_{\delta}\, \mathcal{L}_{\text{cos}}\!\left(f^{l}(x+\delta_t), f^{l}(x)\right)\right\|_2}.
\end{equation}
}




%% file: exp.tex
\begin{figure*}[t]
\begin{center}
\begin{subfigure}{0.95\textwidth}
\includegraphics[width=\textwidth]{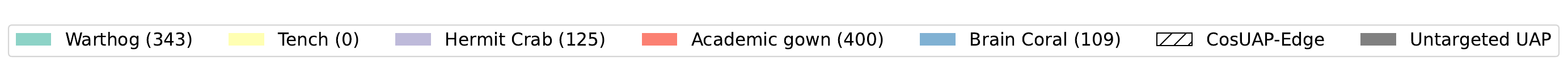}
\end{subfigure}
\begin{subfigure}{0.95\textwidth}
\includegraphics[width=\textwidth]{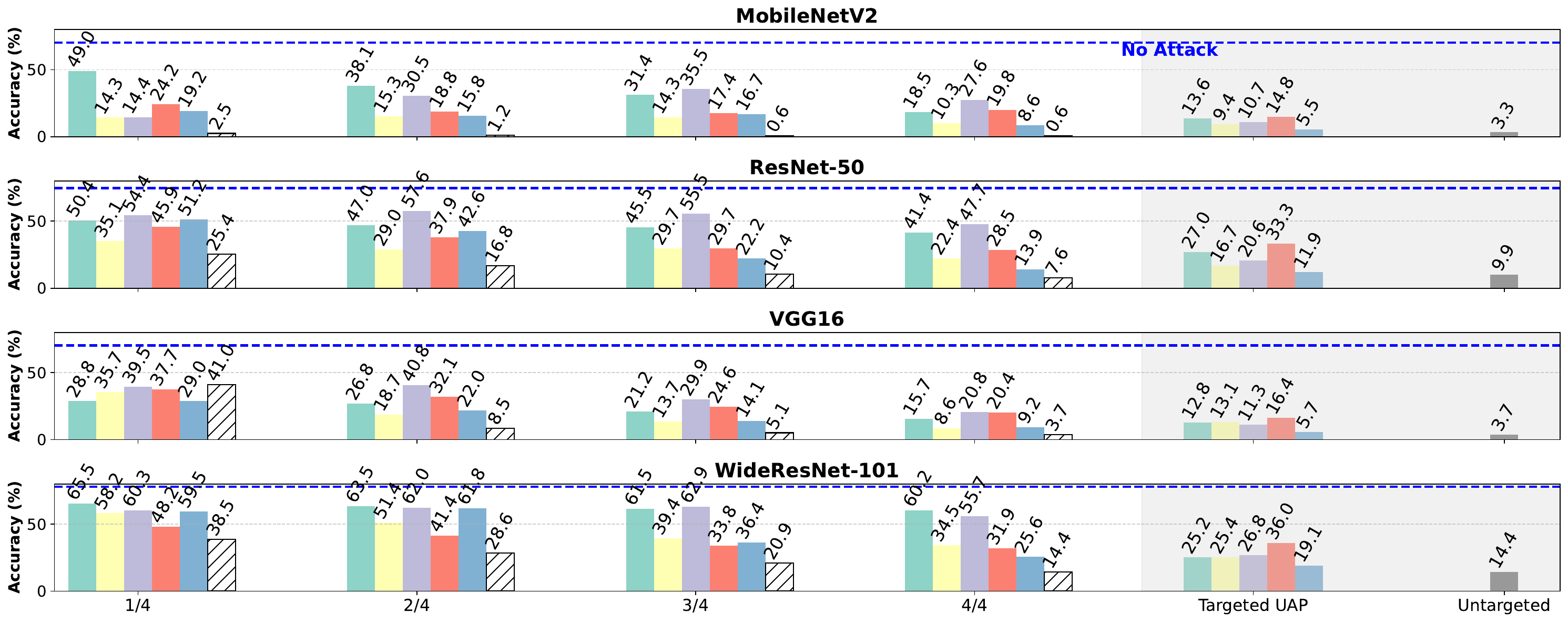}
\caption{$\epsilon = 10/255$}
\end{subfigure}
\begin{subfigure}{0.95\textwidth}
\includegraphics[width=\textwidth]{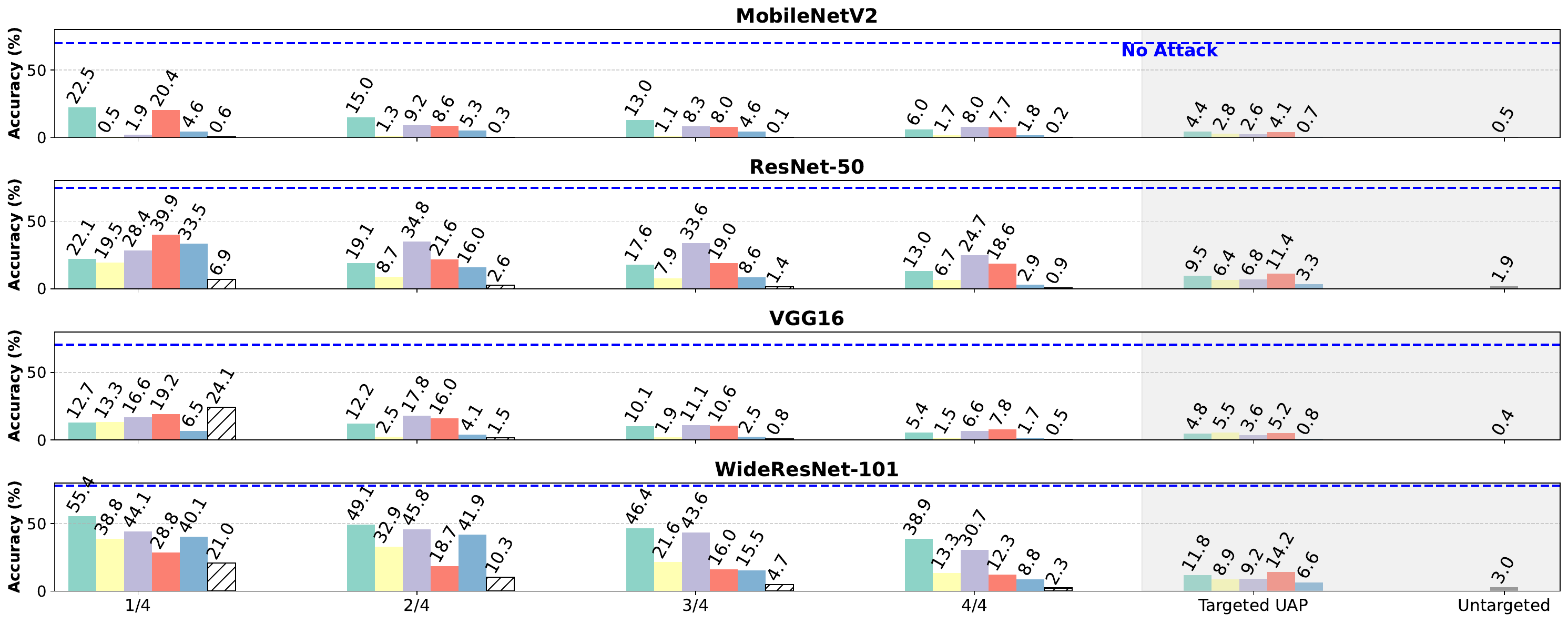}
\caption{$\epsilon = 16/255$}
\end{subfigure}
\caption{
 \small{\add{Models accuracy with edge-only universal adversarial attacks (white part) with $\epsilon=10/255$ (a) and $\epsilon=16/255$ (b) across different edge-sizes. Classic UAP formulation \cite{Zhang_data-free}, both in a targeted and untargeted forms, are shown in the grey area for comparisons.}}}
\label{f:attack_results}
\end{center}
\end{figure*}

\section{Experimental Evaluation}
\label{s:exp}
\add{After introducing the experimental setup, this section presents results for the proposed attacks. The analysis investigates how both the untargeted and targeted formulations are able to induce strong misclassification effects in the cloud-side model, and how their performance compares with classic UAPs. Then, for the targeted setting, we conduct analyses to understand how objectives enforced at the edge-layer representations propagate through the unknown cloud part of the network. Finally, ablation studies and additional experiments were conducted, aimed at validating the effectiveness of the attacks and the threat model proposed.}

\subsection{Experimental Settings}
\label{sec:settings}
All the experiments were conducted on the ImageNet dataset \cite{NIPS2012_c399862d}.
We used class-balanced subsets of the ImageNet validation set: 5000 images as a test set for evaluating all attacks, 1000 images for \( \mathcal{D}_o \), and other 1000 images for \( \mathcal{D}_{\text{opt}} \). While \( \mathcal{D}_t \) was composed of only 50 samples for each selected target class, these also extracted from the validation set.
For the attacked models, we selected CNNs architectures commonly used in distributed scenarios: ResNet50 \cite{he2016deep_resnet}, Wide-ResNet101 \cite{Zagoruyko2016WideRN_wideresnet}, MobileNetV2 \cite{sandler2018mobilenetv2}, and VGG16 \cite{Simonyan2014VeryDC_vgg}\footnote{Weights loaded from \url{torchvision}.}.
Additionally, we provide experimental results with the integration of different datasets for out-of-distribution analysis (CIFAR-10), as well as the use of ViT models to extend the analysis beyond the classic CNNs typically used in distributed settings (Sec. \ref{sec:additional_test}).

To avoid an overly fine-grained analysis of internal layers, where feature representations between adjacent layers may exhibit minimal variation, we selected \emph{four key layers} in each network, representing different depths within the backbone of each model. Specifically, these layers are chosen at $4$ depths of the backbone. Each of these layers serves as a potential split point, defining four configurations of the edge and cloud components, $f_{\textit{edge}}$ and $f_{\textit{cloud}}$, corresponding to different edge depths.
\add{
For clarity, we denote the configuration of the edge component by the depth of the selected key layer that serves as the split point. For example, an edge depth of \(1/4\) means that the edge component includes all layers up to the first major block of the backbone (e.g., the first block in the ResNet architecture). Conversely, an edge depth of \(4/4\) indicates that the edge part covers the entire backbone of the model, with the cloud part consisting only of the classification head. Details about the specific layers selected for each model in this configuration are available in the appendix and in the provided source code.
}
Note also that, given the edge depth, the set $L_{\text{edge}}= \{l_1,..., l_e\}$ (defined in Sec. \ref{s:attack}) is formed by the key layer $l_e$ acting as a split point and the preceding key layers. 

For the targeted attack presented, the binary classifiers \( g^l \) consist of a simple and lightweight module (avoiding overfitting) that includes: a conv block with kernel size 3, followed by an AvgPool and two fully connected layers with hidden size 128, separated by a ReLU activation. More details about the classifiers are provided in the appendix, while related ablation studies are in Sec..\ref{sec:ablation_binary_classifier}.

Regarding the attack settings, we focused on $l_\infty$ norm attacks with different epsilon values, using $\alpha = 2/255$ and the number of epochs for the UAP optimization to $20$, with a batch size of $100$. 
All experiments were conducted using an A100 GPU with 40GB of memory. 

\subsection{Untargeted Effectiveness}
\add{
To analyze the degradation in accuracy induced by the proposed UAPs in both targeted and untargeted settings, Fig.~\ref{f:attack_results} reports the impact on test-set accuracy when applying edge-only UAPs with perturbation magnitudes of \( \epsilon = \frac{10}{255} \) and \( \epsilon = \frac{16}{255} \). 
For the targeted formulation, we randomly selected five target classes from ImageNet and evaluated the effectiveness of edge-only attacks across different depths of the edge portion of the model. For comparison, we also included classical UAP attacks (both targeted and untargeted) based on a cross-entropy loss \cite{AnalysisDezfooli17}. It is important to note that classical UAP methods assume full knowledge of the model and labels.
For the untargeted analysis, we evaluated our extension of the CosUAP attack \cite{Zhang_data-free} in an edge setting, as discussed in Sec.~\ref{eqn:cosuap_edge}.
}

As reasonably expected given the nature of the attack, attacking a shallow edge part, hence basing on limited model knowledge, generally reduces the attack effectiveness (i.e., higher prediction accuracy) compared to cases with access to more layers of the distributed model. This trend is more pronounced for models like ResNet50 and Wide-ResNet101, while different observations arise for MobileNetV2 and VGG16, where edge-only UAPs, even when targeting the initial layers only, achieve similar effectiveness to full-knowledge attacks.
\add{
Additionally, in line with prior work \cite{MoosaviUAPCVPR}, which suggests the existence of dominant labels occupying large regions of the image space, we observed that the choice of the target class has a significant impact on the strength of the resulting UAPs. This effect is especially pronounced in the edge-only setting (e.g., see the classes “warthog” and “crab” in Fig.~\ref{f:attack_results}), where less dominant classes have lower transferability.

Different considerations arise for untargeted approaches, such as CosUAP-Edge (shown in the plot), which tend to induce a larger drop in accuracy even when applied at the first or second edge configurations, without requiring the selection of a highly transferable class. This demonstrates the benefits of CosUAP-Edge attacks; however, they do not provide any control over the specific target class.
}

\begin{figure}[t]
\begin{center}
\centering
\includegraphics[width=0.9\columnwidth]{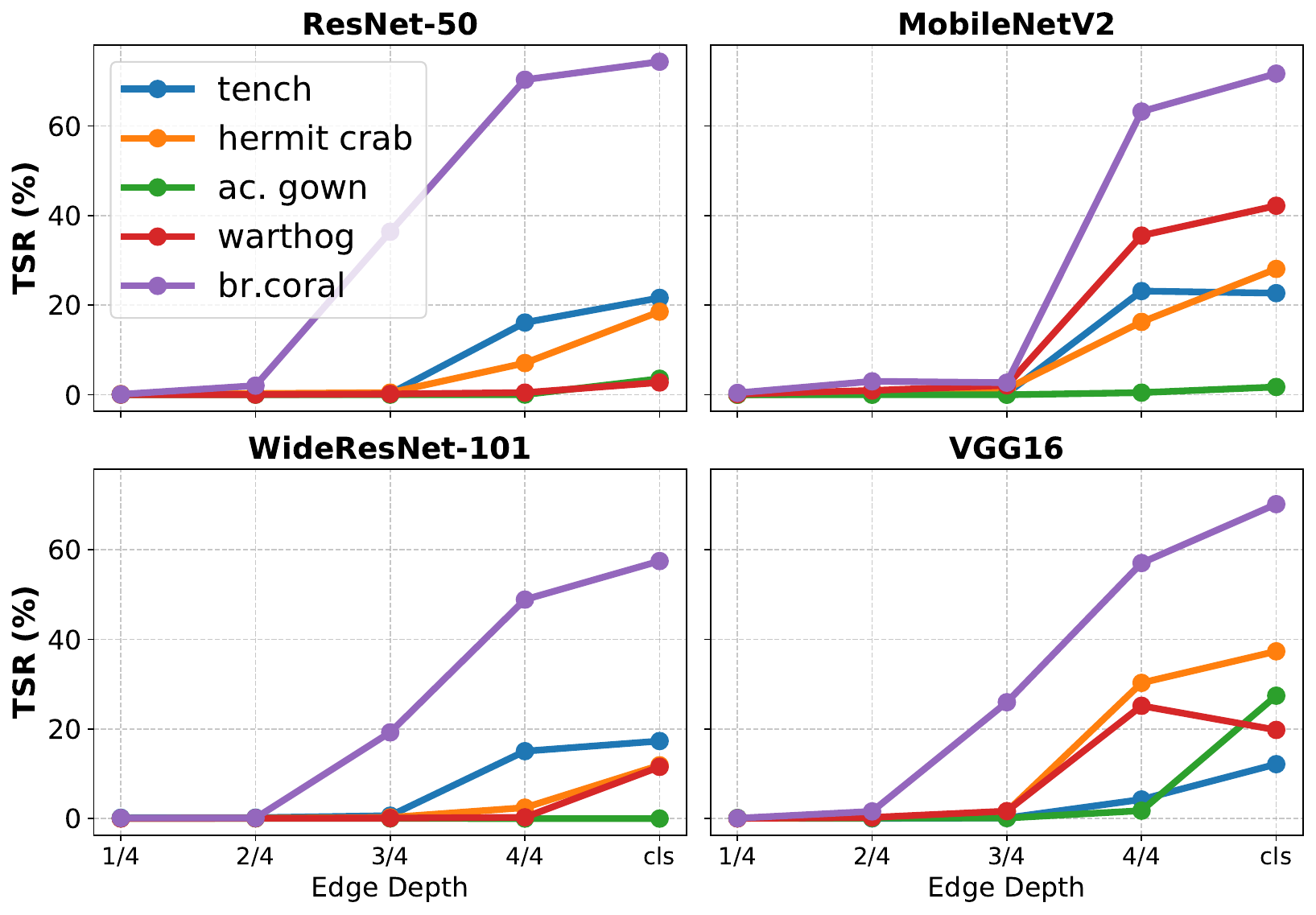}
\caption{\small{Target success rate across attack performed with $\epsilon=10/255$ for different edge depths and different target classes.}}
\label{f:target_success_rate}
\end{center}
\end{figure}

\input{new_tables/tsne_exp_and_feature}

\subsection{Understanding Target Features Across Layers}
While the proposed targeted attacks produce effective mispredictions, we also aim to assess the extent to which cloud outputs align with the intended targeted objective and how much steering the edge features towards a target may impact the remaining decision process of the model.
To evaluate this, Fig. \ref{f:target_success_rate} reports the target success rate (TSR) on the test set (i.e., the percentage of predictions matching the target class) when using edge-only targeted UAPs crafted for various depths of the edge part. From the results, it is clear that \textit{\textbf{(a)} low-depth attacks do not induce mispredictions toward the target class}. In fact, only attacks applied to the last layers of the backbone achieve a higher target success rate, which further increases when the attack is directly applied to the model logits (‘cls’, assuming full model knowledge).

\add{
To gain deeper insights into how much the target features computed in the early layers move the perturbed samples closer to the target class, we plot in Fig. \ref{f:tsne_analysis} a projection of the features extracted by ResNet50's key layers using T-SNE, distinguishing between target and non-targeted training samples. Then, we projected in the same space a clear test sample (yellow point) alongside its perturbed versions (green points), when using edge-only UAPs computed at different edge depths ($1/4$, $2/4$, $3/4$, and $4/4$), as indicated by the labels in the plot. Note that the features extracted at lower depths are less separable, and all the perturbed versions move the original sample in a region closer to the target ones.
However, this trend does not hold while moving in the deeper feature space, as shown in Fig. \ref{f:tsne_analysis} with depths $3/4$ and $4/4$, where early-stage perturbation moves the test sample not sufficiently close to target regions. 
This analysis suggests that \textit{\textbf{(b)} finding the correct direction without knowledge of the remaining processing part of the model could be challenging}, as perturbations computed in early stages cannot guide the sample toward the final target features subspaces.
}

\begin{figure*}[t]
\begin{center}
\begin{subfigure}{0.97\textwidth}
\centering
\includegraphics[width=\textwidth]{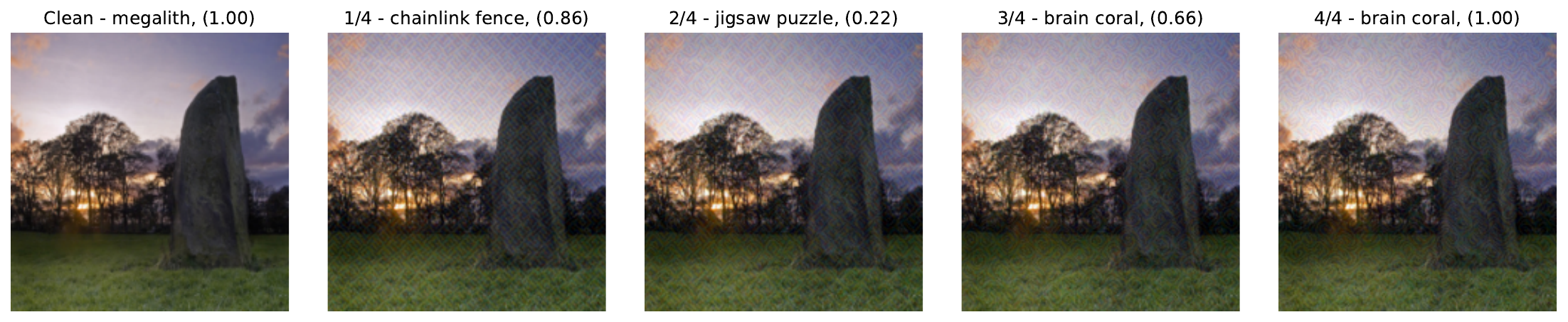}
\end{subfigure}
%
%
%
%
\caption{\small{Attacked images on ResNet50 with edge-only UAPs at different depths (1/4, 2/4, 3/4, 4/4), $\epsilon=10/255$ and target 'brain coral'.
}} 
\label{f:resnet_additional_ill}
\end{center}
\end{figure*}

\input{only_perturbations}

\add{
Finally, even though the proposed UAPs do not fully propagate the targeted effect to the cloud part, we wanted to understand their output effect. For this purpose, we plotted the five most frequent predictions from the test set in Fig.\ref{f:output_distribution_analysis} with MobileNetV2 (top Fig., target 'tench') and ResNet50 (bottom Fig., target 'brain coral'). Each bar plot in the figures shows the results of an attack performed at different edge depths, while the leftmost bars in green denote the no-attack baseline.
As expected, without perturbations, the distribution of the prediction remains fairly balanced (the most frequent class is predicted only about 10 times), consistently with the high clean accuracy of the model. However, this frequency distribution drastically changes  when perturbations are applied (note values on the x-axis), even at early layers (first model depth). Although initial-layer perturbations may not steer the prediction towards the target class, \textit{\textbf{(c)} in the cloud part the attack tends to induce a significant shift in features towards a different reference classes}. 
This insight suggests an interesting direction for future investigation, focusing on potential intra-class relationships that could guide the attack towards a target class by optimizing for a different one.
}

\subsection{Visualization of the Perturbations.}
In Fig. \ref{f:resnet_additional_ill}, we report some visualizations of the attack with $\epsilon = 10/255$, showing perturbed input samples alongside their corresponding predictions and confidence scores for 'brain coral' as the target class, generated from different edge depths.

As shown in the figure, the ability to successfully induce an output prediction consistent with the target class varies depending on the class itself, with some classes being inherently more difficult to reach. Additionally, it is worth noting that perturbations crafted from shallower layers tend to emphasize high-frequency components. This results in more visually perceptible and aggressive perturbations, which is expected, as shallow features are more fine-grained and sensitive to low-level details.
This observation is further illustrated in Fig. \ref{f:perturbations}, where we visualize UAPs generated from four different layers in both ResNet50 and MobileNetV2. 


\subsection{Ablation Studies}
\label{ss:ablation}
\subsubsection{Attack Optimization Steps}
We conducted ablation studies on the proposed attacks to separately assess the steps involved in optimizing edge-only perturbations (Sec. \ref{ss:opt_uap}). Specifically, in Table \ref{table:ablation_results} we reported the model accuracy under edge-only UAPs obtained under different settings. We evaluated the impact of combining the gradient of multiple edge layers during optimization ($\sum_{l\in L_{\text{edge}}}$ in Eq.~\eqref{eqn:phi}) and the effect of gradient normalization (denominator of the last term of Eq.~\eqref{eqn:phi}), referred to as "MultiL" and "Norm" in Table \ref{table:ablation_results}, respectively.
As shown in the table, applying both these approaches results in improved attack effectiveness. The use of multiple layers helps reduce overfitting in the feature separation analysis within deeper layers, while normalization helps maintain balanced gradients across all layers.

\begin{table}[t]
    \centering
    \resizebox{0.95\columnwidth}{!}{%
    \begin{tabular}{c|c|c|c|c|c}
        \hline
        \multicolumn{6}{c}{\textbf{ResNet50}} \\
        \hline
         &  & \multicolumn{4}{c}{\textbf{Edge Depth}} \\
        \hline
        \textbf{Norm} & \textbf{MultiL} & \textbf{1/4} & \textbf{2/4} & \textbf{3/4} & \textbf{4/4} \\
        \hline
         & & 37.01 & 30.21 & 35.33 & 25.92 \\
        \hline
        \checkmark &  & 33.13 & 23.42 & 34.43 & 26.66  \\
        \hline
         & \checkmark  & 37.01 & 36.29 & 31.11 & 24.52 \\
        \hline
        \checkmark & \checkmark & \textbf{33.13} & \textbf{28.96} & \textbf{29.68} & \textbf{22.42}  \\
        \hline
    \end{tabular}%
    }
    \resizebox{0.95\columnwidth}{!}{%
    \begin{tabular}{c|c|c|c|c|c}
        \multicolumn{6}{c}{\textbf{MobileNetV2}} \\
        \hline
         &  & \multicolumn{4}{c}{\textbf{Edge Depth}} \\
        \hline
        \textbf{Norm} & \textbf{MultiL} & \textbf{1/4} & \textbf{2/4} & \textbf{3/4} & \textbf{4/4} \\
        \hline
         & & 16.77 & 18.41 & 30.23 & 19.51 \\
        \hline
        \checkmark &  & 12.25 & 18.45 & 18.23 & 19.57  \\
        \hline
         & \checkmark  & 16.77 & 18.23 & 27.57 & 18.95 \\
        \hline
        \checkmark & \checkmark & \textbf{12.25} & \textbf{15.29} & \textbf{14.33} & \textbf{10.25}  \\
        \hline
    \end{tabular}%
    }
    \caption{\small Ablation studies on the impact (in terms of accuracy on ImageNet) of multiple edge layers' gradients and gradient normalization. Attacks were performed on MobileNetV2 and ResNet50, with $\epsilon = 10/255$ and 'tench' as the target class.}
    \label{table:ablation_results}
\end{table}

\subsubsection{Varying the number of samples}
\begin{figure}[t]
\begin{center}
\begin{subfigure}{0.49\columnwidth}
\includegraphics[width=\textwidth]{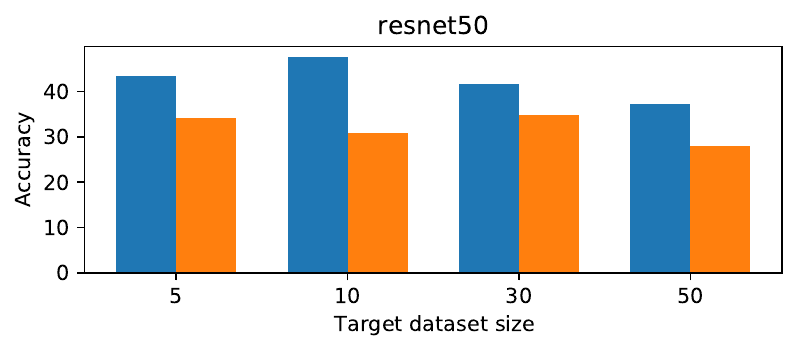}
\end{subfigure}
\begin{subfigure}{0.49\columnwidth}
\includegraphics[width=\textwidth]{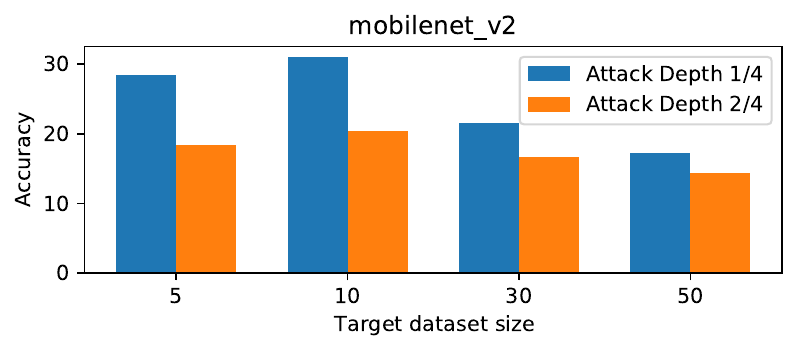}
\end{subfigure}
\begin{subfigure}{0.49\columnwidth}
\includegraphics[width=\textwidth]{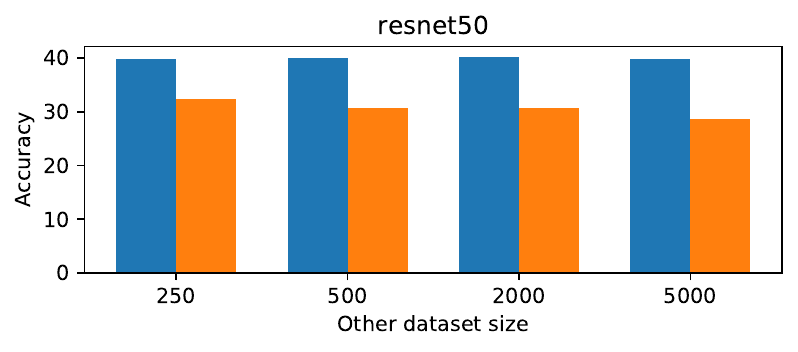}
\end{subfigure}
\begin{subfigure}{0.49\columnwidth}
\includegraphics[width=\textwidth]{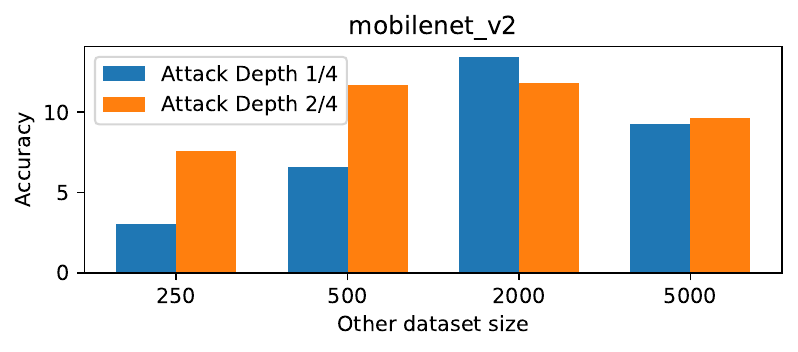}
\end{subfigure}
\caption{\small{Analysis of the accuracy across different edge-only attacks (edge having 1/4 and 2/4 sizes of the backbone) when varying the number of target samples (top part) and the number of other samples (bottom part) used to learn feature separations. The attack was conducted with target class 0 ('tench') and $\epsilon = 10/255$.}}
\label{f:target_samples_analysis}
\end{center}
\end{figure}

We conducted additional analysis to better understand the variation in attack effectiveness when changing the number of samples used to train the binary classifiers, while keeping the number of optimization iterations fixed. 
In particular, Fig. \ref{f:target_samples_analysis} shows an analysis where the number of samples in the "other dataset" (non-target classes) is fixed at 1000, while the number of target-class samples varies. 
As shown in the plot, results on ResNet50 and MobileNetV2 indicate that, in general, having more target samples can be beneficial, as the classifier learns to generalize better across different classes.

A different trend emerges when varying the number of other-samples (non-target samples), as shown in Fig. \ref{f:target_samples_analysis} (bottom), where the number of target samples is fixed at 50. No notable change was observed with ResNet50, where the results appear approximately stable, suggesting that varying the number of non-target samples does not significantly affect model performance. However, MobileNetV2 shows a different behavior, with fewer non-target samples proving more beneficial for the attack. We believe that intrinsic properties related to the selection of non-target samples could impact specific models, and we hope to investigate possible sample relationships and intriguing aspects in more detail in the future.

\subsubsection{On the size of the binary classifiers} 
\label{sec:ablation_binary_classifier}
\begin{figure}[t]
  \centering
  \includegraphics[width=1.\linewidth]{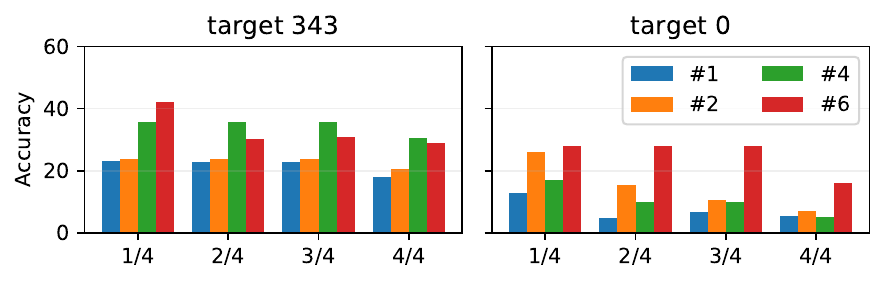}
   \caption{\small{Analysis of the complexity of binary classifiers whit ResNet50 on ImageNet for different target classes (343 and 0). The bars compare the number of convolutional blocks used to perform edge-side attacks from different edge depths.}}
   \label{fig:classifiers_complexity}
\end{figure}

We conduct ablation studies to assess the impact of selecting simple or complex binary classifiers for learning feature separation in the edge space. Specifically, in Fig. \ref{fig:classifiers_complexity}, we illustrate the accuracy reduction observed when performing the proposed attacks at two different edge stages while varying the depth of the binary classifier, ranging from a single convolutional block (\#1), as in the standard setting, to four convolutional blocks (\#4). Please see the Appendix for additional architecture details.
As shown in the bar plots, while increasing the number of convolutional blocks may improve feature separation, this also shifts the binary classifier toward further processing the features rather than preserving the separation learned by the edge part of the model. As a result, the attack's effectiveness decreases when leveraging the edge model's gradients. Therefore, to enhance transferability to the cloud part, a simpler configuration proves to be more effective.

\subsection{Additional test} 
\label{sec:additional_test}
\subsubsection{Variation of the success rate with $\epsilon$} 
\begin{figure}[t]
\begin{center}
\includegraphics[width=0.95\columnwidth]{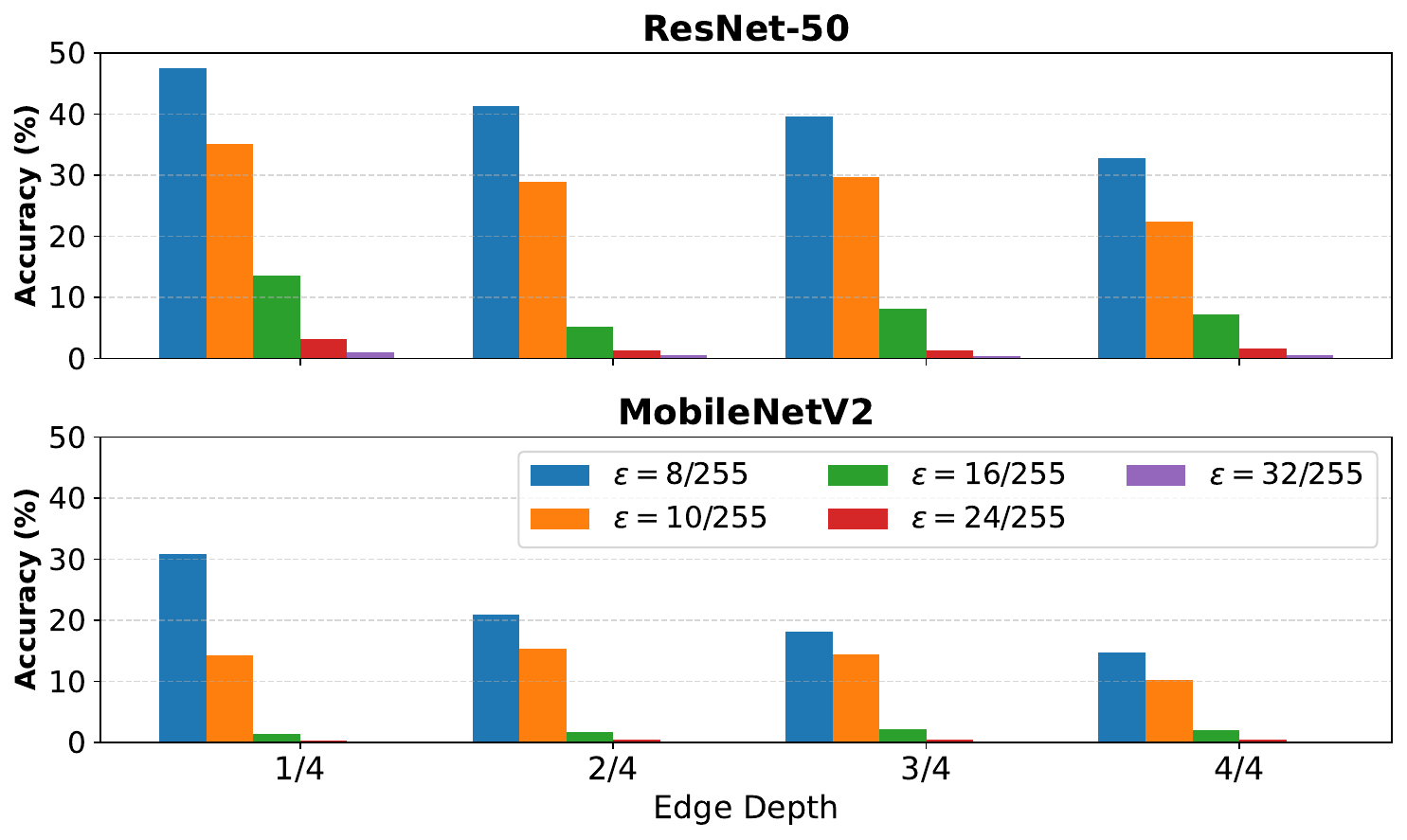}
\caption{\small{Analysis of the effectiveness of edge-only universal perturbation whit different $\epsilon$-magnitude on ResNet50 and MobileNetV2.}}
\label{f:eps_analysis}
\end{center}
\end{figure}

In Fig. \ref{f:eps_analysis}, we reported the model accuracy when varying the attack magnitude \( \epsilon \) for edge-only UAPs across different layers. For this test, we considered the class 'tench' as the target reference class for both ResNet50 and MobileNetV2.
As illustrated in the plot, the effectiveness of the attack varies significantly, while \( \epsilon = \frac{16}{255} \) appears to be a reasonable compromise for high effectiveness and low magnitude. A clarification of the magnitude of the features of the edge-side attacks is shown in  Fig. \ref{f:perturbed_images}, which includes visual examples of perturbed images generated from the first layers (edge depth 1/4) of MobileNetV2, targeting mostly high-frequency components, as shown in Fig. \ref{f:perturbations}, due to the nature of early network features.

\input{images_perturbations}
\subsubsection{Analysis of the out-of-distribution attack dataset} 
In the experimental settings studied in the main experimental part, we evaluated the proposed attack using images from the validation set of ImageNet for both the target and other-samples optimization datasets. Although these images were not used to train the tested model, their distribution may share some similarities, such as resolution and source characteristics, given that they originate from the same dataset (despite the vast diversity of ImageNet).
To better investigate the impact of using completely different images, we considered in Table \ref{tab:ood_test} a dataset with a distinct distribution for the other-set samples used to learn feature separation from non-target classes. Specifically, we used the CIFAR-10 test set to form the other-sample set while retaining only the 50 target images from the ImageNet validation set. As shown in Table \ref{tab:ood_test}, the attack remains effective in reducing accuracy even when using entirely different images in the other-sample set. However, a decrease in effectiveness is observed, likely due to the higher distributional differences between CIFAR-10 images and the high-resolution images learned by the model. Nevertheless, the binary classifier still detects separability from the target samples, leading to effective untargeted attacks on the model's output. Differences with respect to Fig. \ref{f:attack_results} for $\epsilon = 16/255$ are shown in parentheses in Table \ref{tab:ood_test}.
\begin{table}[t]
\centering
\resizebox{1.\columnwidth}{!}{%
\begin{tabular}{l|rrrr}
\toprule
{} &  Depth-1 &  Depth-2 &  Depth-3 &  Depth-4 \\
\midrule
MNV2 &     1.70 (+1.24) &     3.12 (+1.86) &     4.24 (+3.14) &     5.49 (+3.77)  \\
VGG16       &    10.39 (-2.88) &     7.33 (+4.83) &     8.31 (+6.41) &     8.07 (+6.59)  \\
RN50    &    21.56 (+2.06) &    12.01 (+3.26) &    17.02 (+9.16) &    16.12 (+9.41) \\
\bottomrule
\end{tabular}
}
\caption{\small{Evaluation using the $\mathcal{D}_o$ set with a completely different distribution (CIFAR-10). The values represent the accuracy on ImageNet, while the values in parentheses indicate the difference compared to the results obtained in the standard experimental settings shown in Figure \ref{f:attack_results}, with $\epsilon$=16/255 and target class 0.}}
\label{tab:ood_test}
\end{table}

\subsubsection{Testing Edge-only UAPs on ViT}
We also evaluate the effectiveness of the proposed attacks across different architectures that, while not frequently addressed in the IoT literature, are still relevant for understanding the attack's impact and exploring its applicability in future scenarios. Specifically, we considered transformer-based models, such as Vision Transformers (ViTs) \cite{dosovitskiy2020image}, in both Base (ViT-B) and Small (ViT-S) versions, applying the same analysis as in previous settings across different layers of the model (Blocks 1, 4, 6, and 10, considering that both ViT-S and ViT-B have a total of 12 blocks).
In this case, the input to the binary classifier is the concatenated multi-head output of the selected blocks, which is directly processed by the fully connected layers without passing through a convolutional part (see Appendix for architecture details).
As shown in Table \ref{tab:vit_results}, the attack remains effective even when applied to transformer models, significantly reducing classification performance despite having access to only a limited portion of the model, which is assumed to be on the edge side.

\begin{table}[t]
    \centering
    \resizebox{1.\columnwidth}{!}{%
    \begin{tabular}{l|c|c|c|c|c|c}
        \hline
        Model & Class & Rand & 1/4 & 2/4 & {3/4} & {4/4}   \\ 
        \hline
        \multirow{3}{*}{ViT-S} 
        & 0 & 71.6  & 40.24 & 27.30 & 19.28 & 12.67 \\ 
        & 109 & - & 36.52 & 37.34 & 25.39 & 14.61  \\ 
        & 343 & - & 47.78 & 41.90 & 29.67 & 23.32 \\ 
        \hline
        \multirow{3}{*}{ViT-B} 
        & 0  & 76.1 & 55.40 & 52.95 & 44.60 & 32.41 \\ 
        & 109 & - & 53.13 & 47.47 & 30.81 & 16.21 \\ 
        & 343 & - & 56.68 & 55.72 & 51.57 & 42.12 \\ 
        \hline
    \end{tabular}
    }
    \caption{\small{Accuracy on ImageNet for ViT-S and ViT-B under edge-side attacks with $\epsilon=16/255$, considering different depths and target classes ('tench'-0, 'brain coral'-109, 'warthog'-343). For comparison, accuracy with respect to random $\epsilon$ noise is also reported.}}
    \label{tab:vit_results}
\end{table}

\input{new_tables/tables_threat_model}
\input{new_tables/table_adversarial_training}

\subsection{\add{Threat Models Comparisons}}
\add{
To better understand the impact of the presented threat model, we provide a comparison with a black-box adversarial setting in Table \ref{tbl:threat_model_comparisons}, where, differently from our assumption, the attacker does not have access to the edge part of the target model but can leverage a locally available architecture that has been trained on a similar data distribution, hoping to achieve high transferability to the unknown target model. 
In the following, we distinguish the analysis of these comparisons between untargeted and targeted attacks.
}

\add{
\medskip
\noindent \textit{Untargeted attack comparison.}
For the untargeted setting (left part in Table \ref{tbl:threat_model_comparisons}, we compare the two threat models by using, for the Edge-Only scenario, the CosUAP attack in the edge setting (Sec. \ref{ss:untarget_attack}), computed across different edge depths. For the black-box scenario, we consider the UAP attack \cite{MoosaviUAPCVPR}, testing different surrogate models (red cells indicate cases where the surrogate and target models are the same, i.e., the ideal white-box setting). From the results, CosUAP-Edge attacks are often more effective than black-box transferability, showing that, in split-inference scenarios where accessing the edge is possible, this provides a more reliable attack.
}

\add{
\medskip
\noindent \textit{Targeted attack comparison.}
For the targeted analysis (right part in Table \ref{tbl:threat_model_comparisons}), extending the study in Fig. \ref{f:target_success_rate}, we present a direct comparison of the targeted approach proposed in Sec. \ref{ss:opt_uap} with the transferability of UAPs computed in a targeted setting (the target class considered is “brain coral”). In this case, the benefits of EdgeUAP are less evident and strongly depend on the available depth. In black-box scenarios, a high targeted success rate is mainly due to the architectural similarity between the surrogate and target models (e.g., ResNet and WideResNet). However, targeted attacks in EdgeUAP can be comparable to strong black-box attacks when the depth of the edge model is sufficiently large (e.g., $3/4$ or $4/4$).
}



\add{
\subsection{Analysis and discussion of defense mechanisms}
\label{ss:defense}
Beyond demonstrating the feasibility and effectiveness of the proposed attacks, it is important to consider potential defense mechanisms and their limitations in the presented setting. Several works have proposed defenses for split and collaborative inference, including feature obfuscation, noise injection, pruning, and representation compression, mainly to mitigate privacy leakage and reconstruction or inversion attacks \cite{defense_privac_9346367, defense_privacy_JIANG2024102728, defense_privacy_survey}. However, these defenses are typically not designed to address adversarial robustness. In particular, most existing approaches focus on protecting data confidentiality rather than preventing the purposeful manipulation of intermediate representations by evasion attacks. This highlights the need for novel defense strategies explicitly tailored to edge-accessible adversarial threat models.

A widely adopted mechanism for improving the robustness of generic neural networks against adversarial examples is adversarial training, which has also been extended to the domain of universal perturbations \cite{shafahi2020universal}. To evaluate the impact of our attack in this setting, we trained models from scratch on the CIFAR-10 training set using an adversarial training regime based on untargeted UAPs with perturbation magnitude $\epsilon = 8/255$, aiming to balance clean accuracy and robustness to universal attacks \cite{shafahi2020universal}. We then evaluated the models on the CIFAR-10 test set and launched edge-only attacks at inference time using $\epsilon = 16/255$.
As shown in Table~\ref{tb:adversarial_training}, while adversarial training improves the overall robustness of the model, our edge-only attack, evaluated in the targeted setting across all CIFAR-10 classes, often gets a close effectiveness of white-box UAPs (e.g., for layers 2, 3, and 4 when targeting class 6), and in some cases even surpasses it. Ideally, a robust model should exhibit increased resistance in the early layers, so that adversarial effects do not propagate to deeper representations. However, results in Table~\ref{tb:adversarial_training} show that perturbations crafted using only edge-layer information can still strongly influence downstream predictions.
These findings suggest that standard adversarial training could be further adapted and explored to address and mitigate edge-side adversarial threats.
}

%% file: new_tables/tsne_exp_and_feature.tex
\begin{figure*}[ht]
\begin{center}
\begin{subfigure}{0.9\textwidth}
\centering
\begin{subfigure}{0.24\textwidth}
\includegraphics[width=\textwidth]{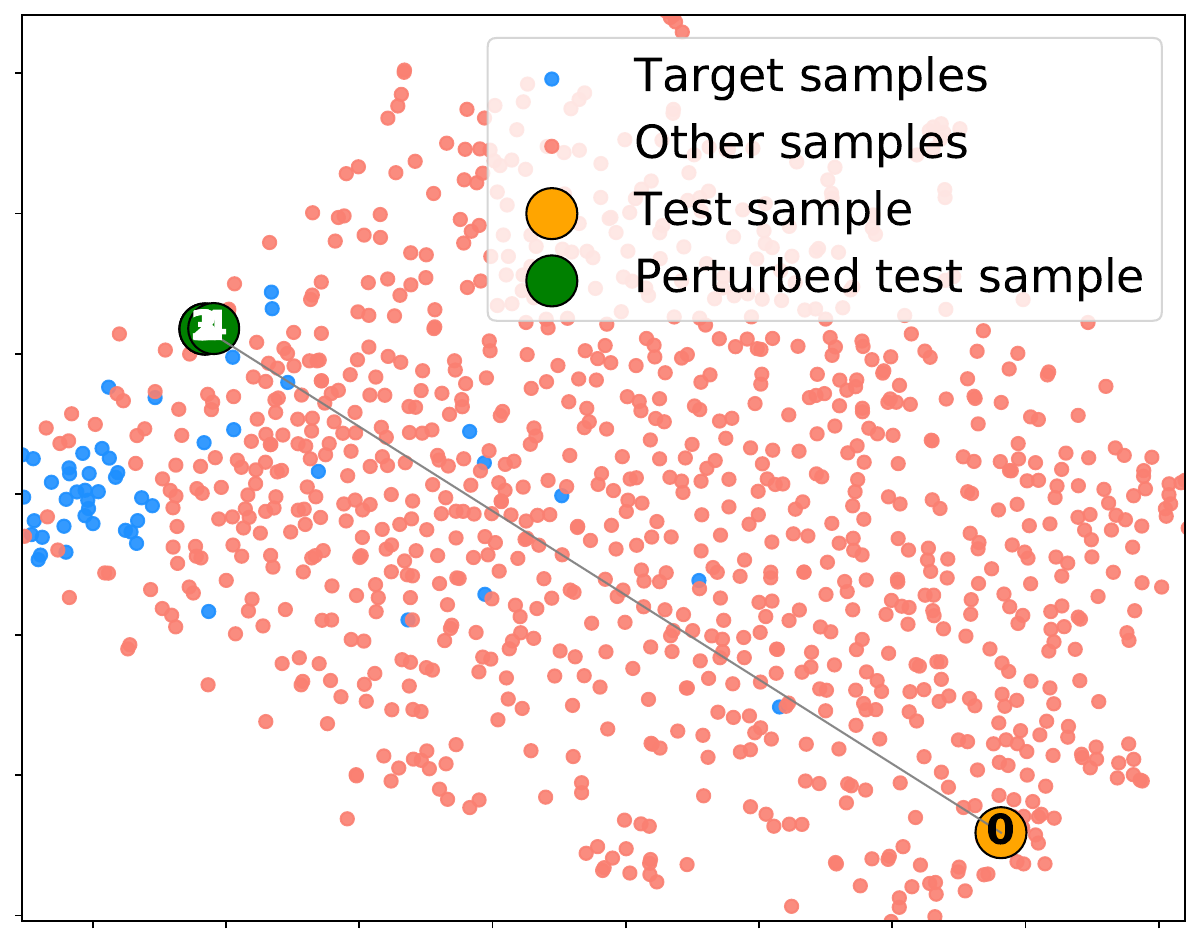}
\caption*{depth $1/4$}
\end{subfigure}
\begin{subfigure}{0.24\textwidth}
\centering
\includegraphics[width=\textwidth]{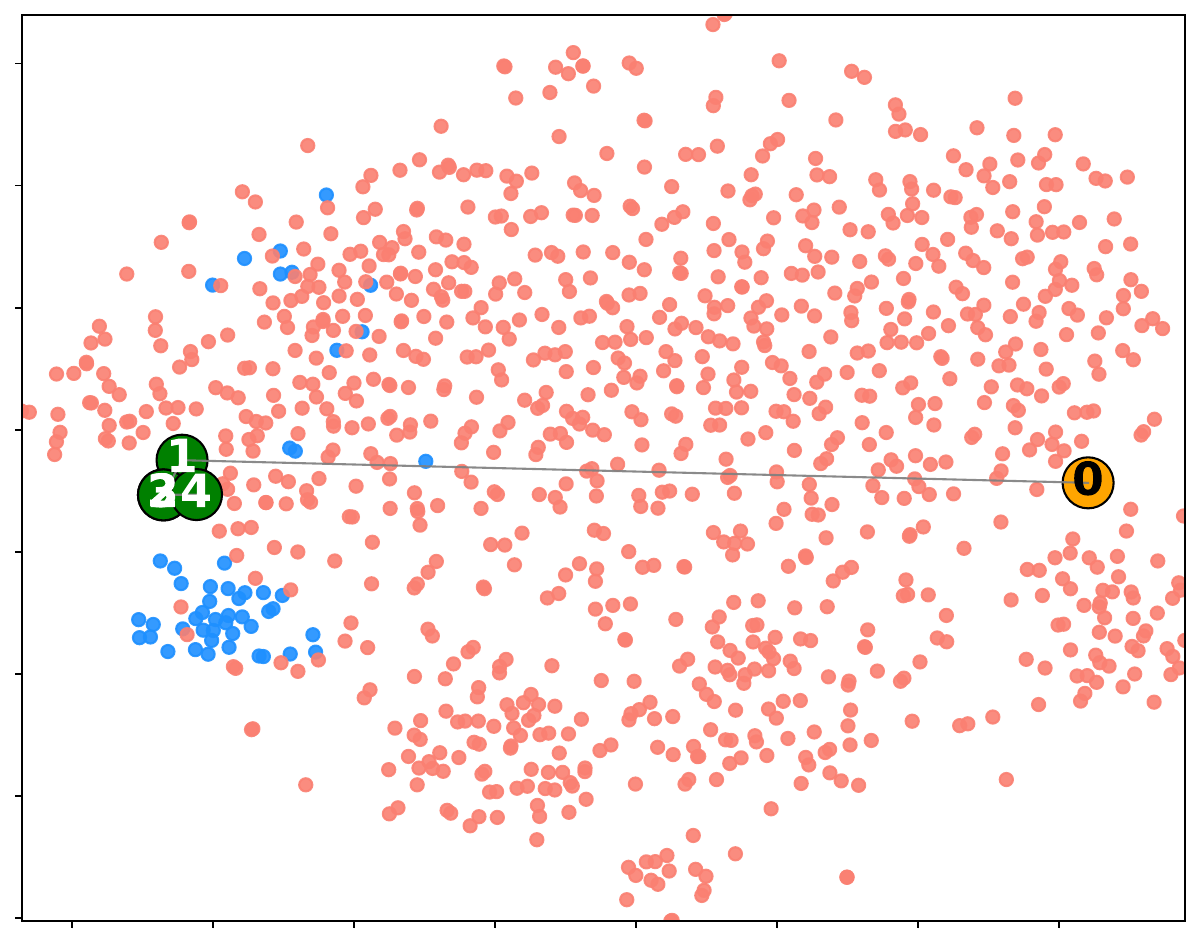}
\caption*{depth $2/4$}
\end{subfigure}
\begin{subfigure}{0.24\textwidth}
\centering
\includegraphics[width=\textwidth]{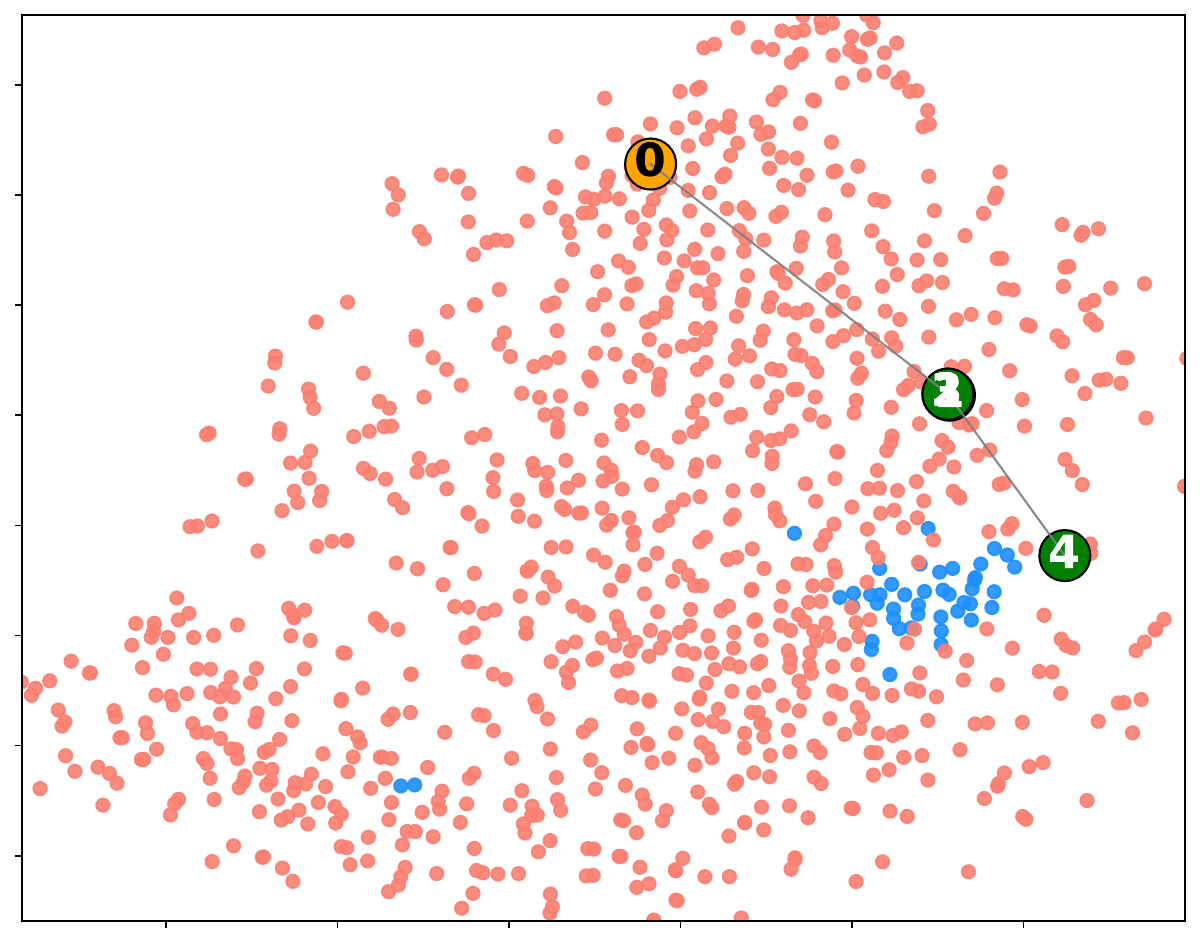}
\caption*{depth $3/4$}
\end{subfigure}
\begin{subfigure}{0.24\textwidth}
\centering
\includegraphics[width=\textwidth]{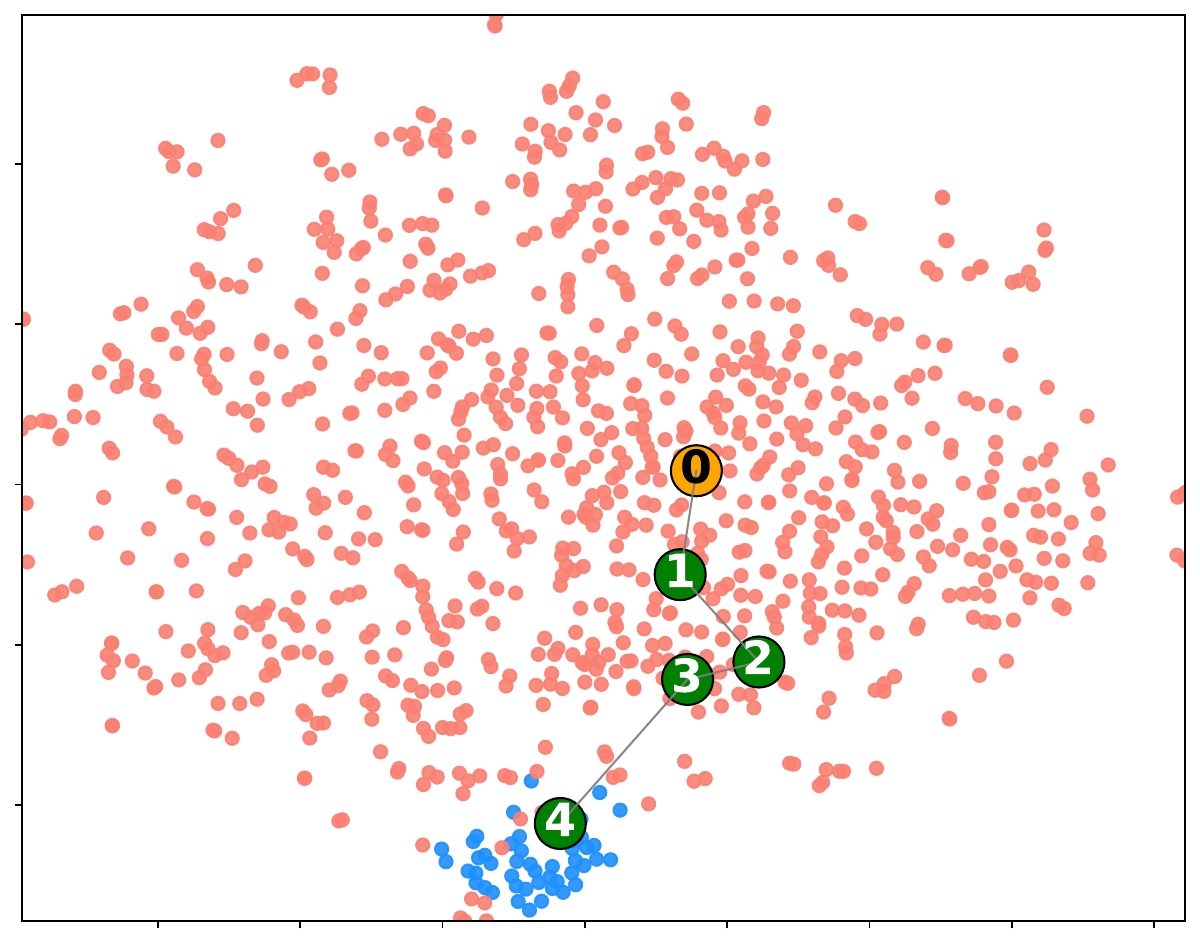}
\caption*{depth $4/4$}
\end{subfigure}
\end{subfigure}

\begin{subfigure}{0.9\textwidth}
\centering
\begin{subfigure}{0.24\textwidth}
\includegraphics[width=\textwidth]{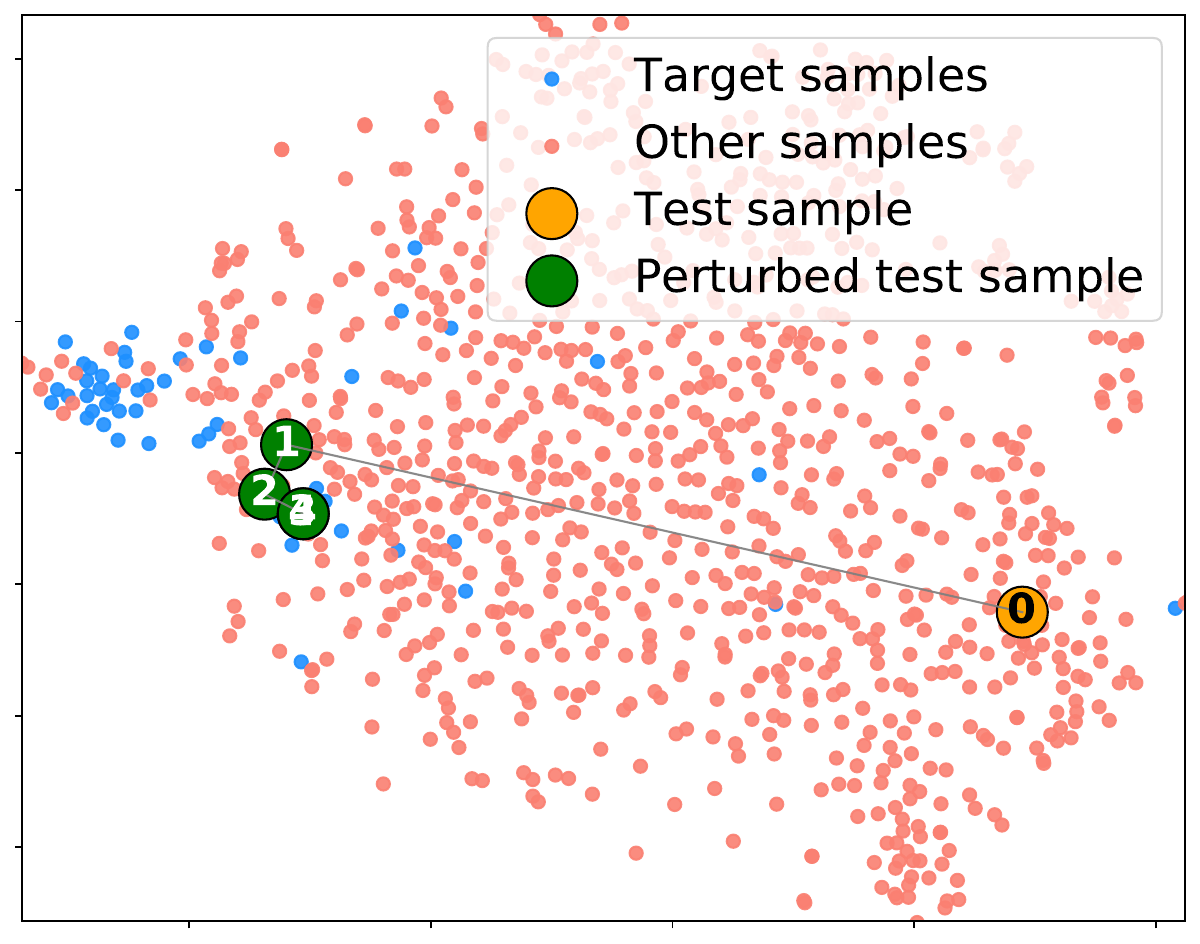}
\caption*{depth $1/4$}
\end{subfigure}
\begin{subfigure}{0.24\textwidth}
\centering
\includegraphics[width=\textwidth]{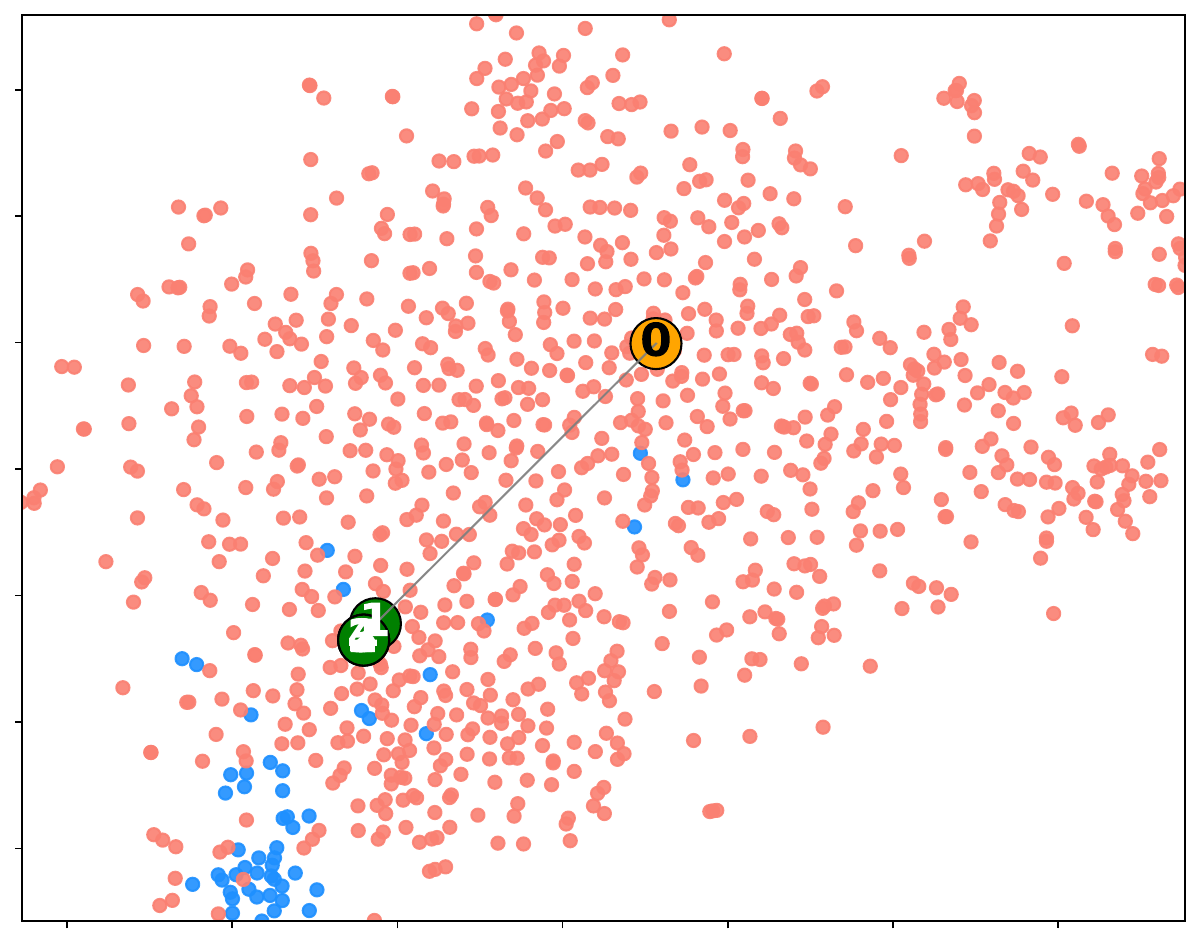}
\caption*{depth $2/4$}
\end{subfigure}
\begin{subfigure}{0.24\textwidth}
\centering
\includegraphics[width=\textwidth]{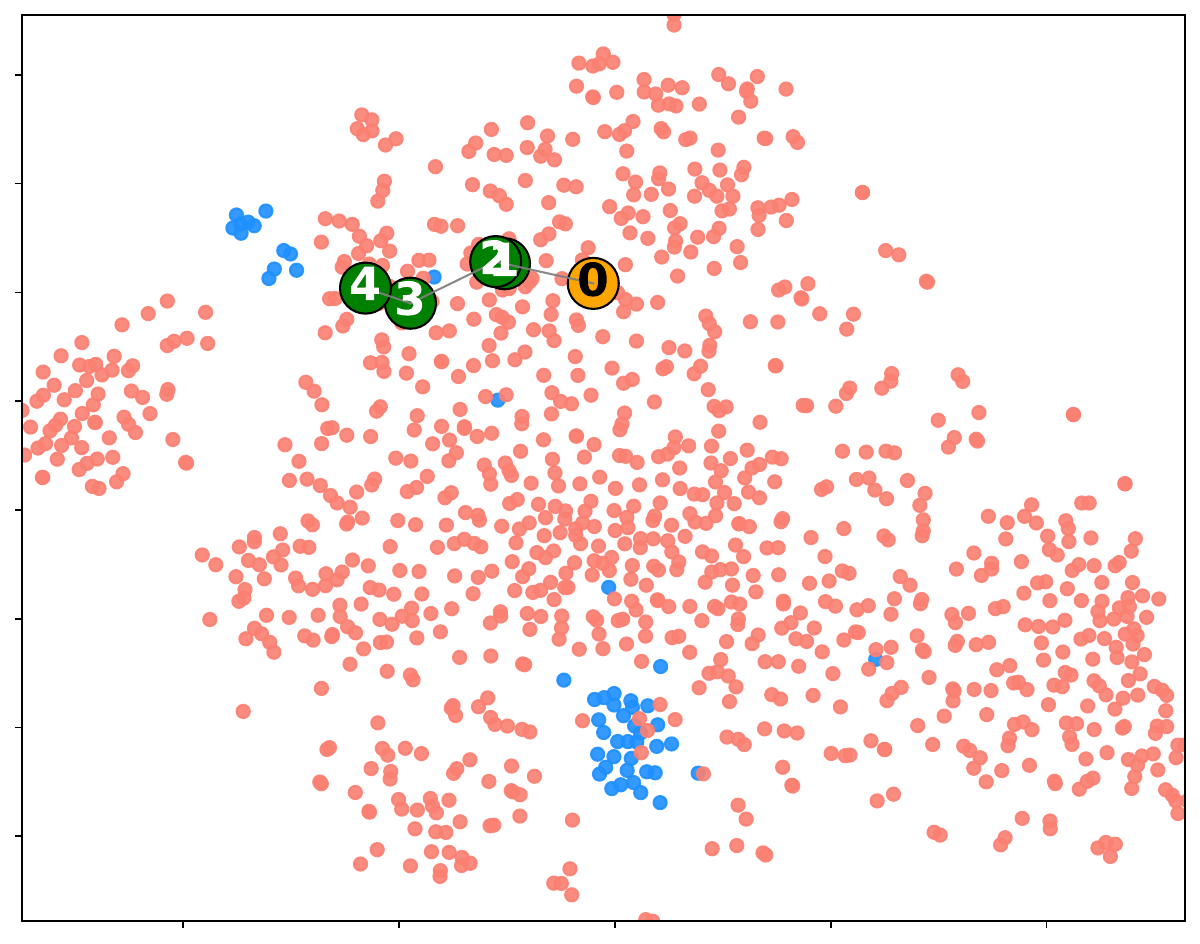}
\caption*{depth $3/4$}
\end{subfigure}
\begin{subfigure}{0.24\textwidth}
\centering
\includegraphics[width=\textwidth]{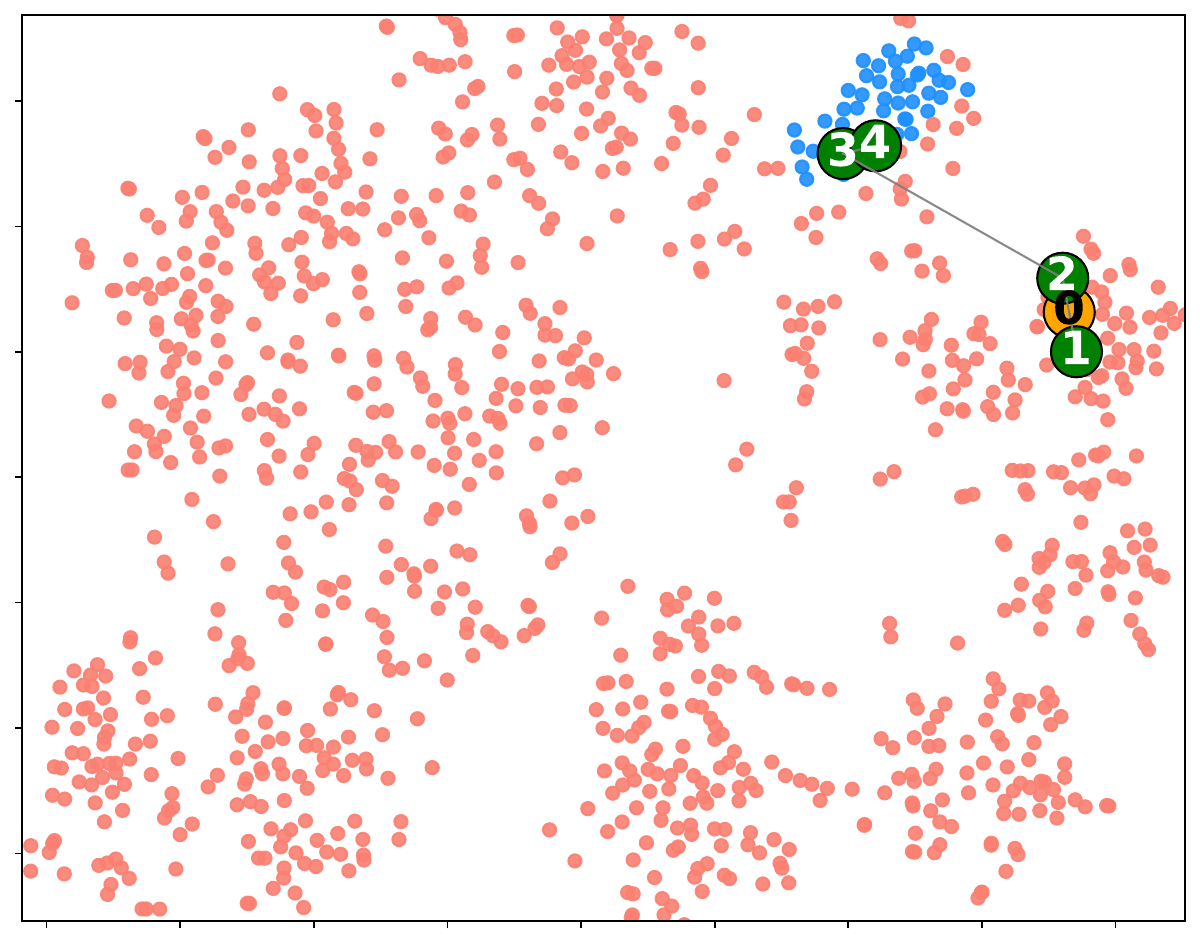}
\caption*{depth $4/4$}
\end{subfigure}
\end{subfigure}
\caption{\small{\add{Impact of the targeted edge-only UAP ($\epsilon = 10/255$) in the layers’embedding space with t-SNE . Visualizations of MobileNetV2 (top) and ResNet50 (bottom) features show 50 samples from the target class (“brain coral,” blue) and 1000 samples from other classes (red). A new test sample (yellow) and its edge-only targeted perturbed versions (green), generated at different depths (numbers).}}}
\label{f:tsne_analysis}
\end{center}
\end{figure*}

\begin{figure*}
\begin{center}
\begin{subfigure}{1.0\textwidth}
\centering
\includegraphics[width=0.95\textwidth]
{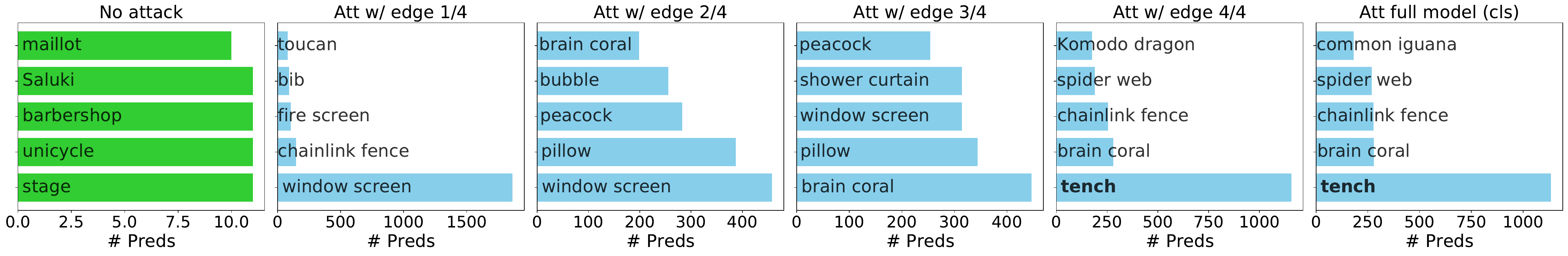}
\includegraphics[width=0.95\textwidth]{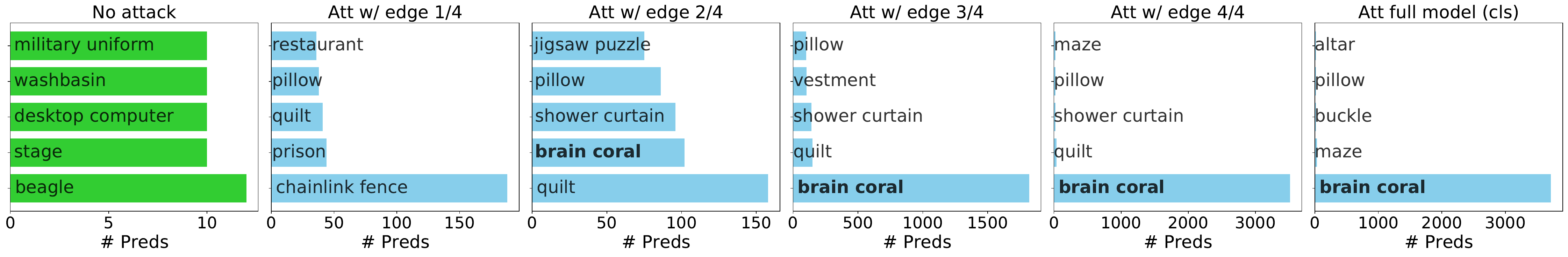}
\label{f:labels_analysis}
\end{subfigure}
\caption{\small{\add{Class prediction frequency is shown across 5000 samples, comparing no attack (green) to edge-only attacks at different depths. The first row shows MobileNetV2 results (target: 'tench') and the second row shows ResNet50 (target: 'brain coral').}}}
\label{f:output_distribution_analysis}
\end{center}
\end{figure*}

%% file: only_perturbations.tex
\begin{figure}[ht]
\centering
\begin{subfigure}{0.95\columnwidth}
\centering
\begin{subfigure}{0.24\columnwidth}
\includegraphics[width=\textwidth]{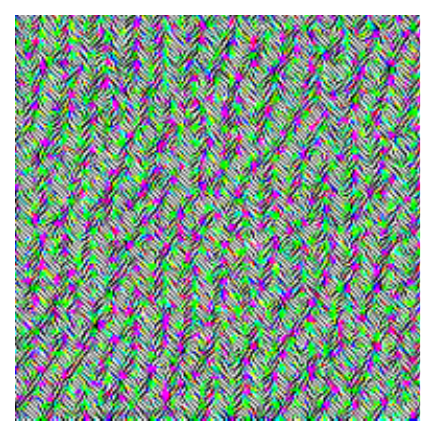}
\end{subfigure}
\begin{subfigure}{0.24\columnwidth}
\includegraphics[width=\textwidth]{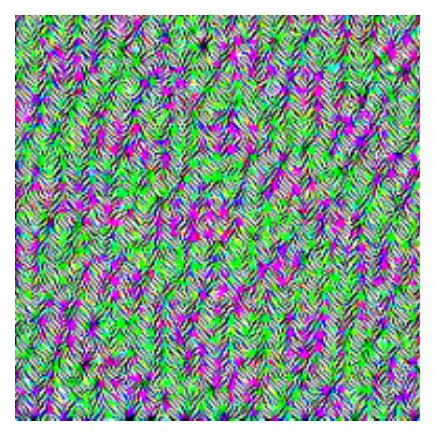}
\end{subfigure}
\begin{subfigure}{0.24\columnwidth}
\includegraphics[width=\textwidth]{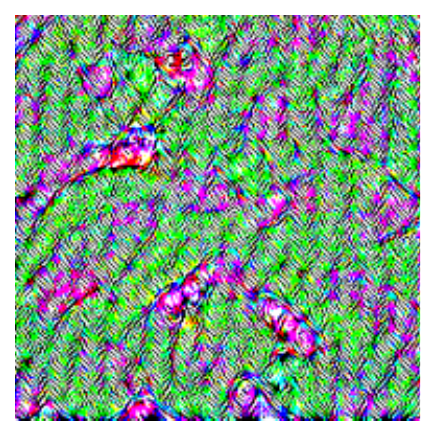}
\end{subfigure}
\begin{subfigure}{0.24\columnwidth}
\includegraphics[width=\textwidth]{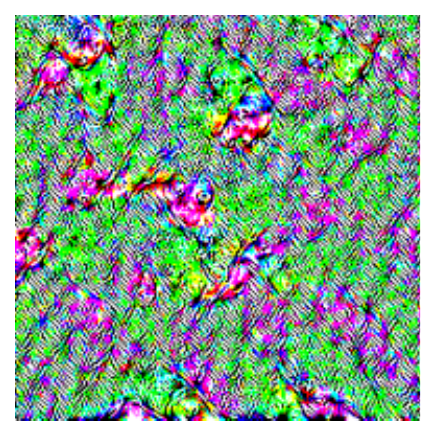}
\end{subfigure}
\end{subfigure}
\begin{subfigure}{0.95\columnwidth}
\centering
\begin{subfigure}{0.24\columnwidth}
\includegraphics[width=\textwidth]{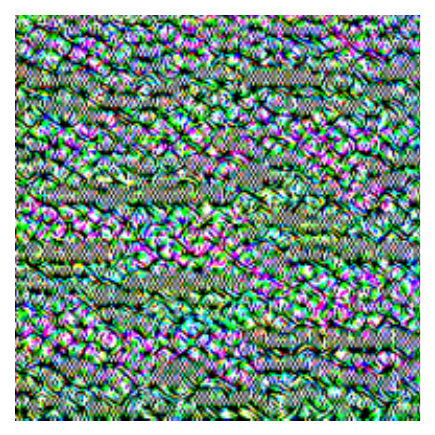}
\caption{depth $1/4$}
\end{subfigure}
\begin{subfigure}{0.24\columnwidth}
\includegraphics[width=\textwidth]{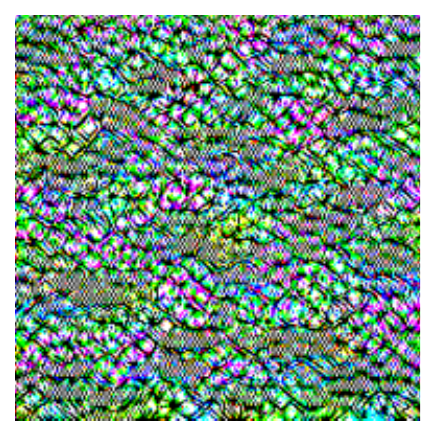}
\caption{depth $2/4$}
\end{subfigure}
\begin{subfigure}{0.24\columnwidth}
\includegraphics[width=\textwidth]{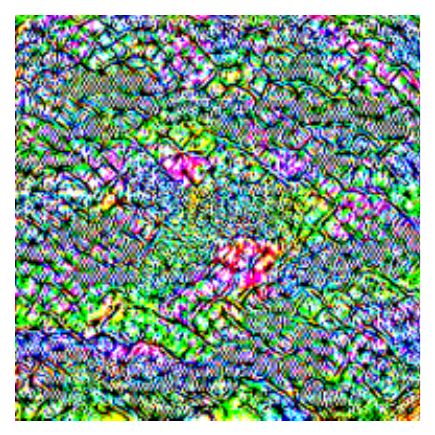}
\caption{depth $3/4$}
\end{subfigure}
\begin{subfigure}{0.24\columnwidth}
\includegraphics[width=\textwidth]{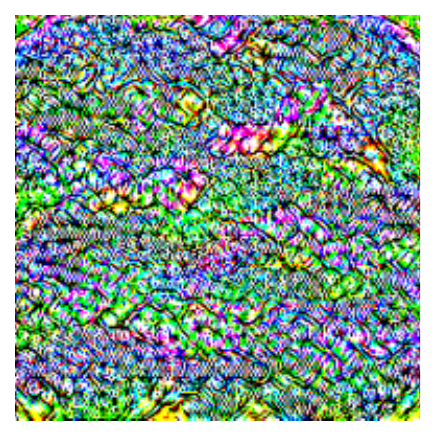}
\caption{depth $4/4$}
\end{subfigure}
\end{subfigure}
\caption{\small{Edge-only UAPs for different edge sizes of ResNet50 (top) and MobileNetV2 (bottom).}}
\label{f:perturbations}
\end{figure}

%% file: images_perturbations.tex
\begin{figure}[ht]
\centering
\begin{subfigure}{0.98\columnwidth}
\centering
\begin{subfigure}{0.30\columnwidth}
\includegraphics[width=\textwidth]{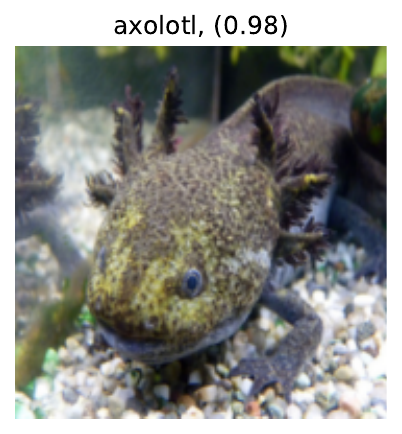}
\end{subfigure}
\begin{subfigure}{0.30\columnwidth}
\includegraphics[width=\textwidth]{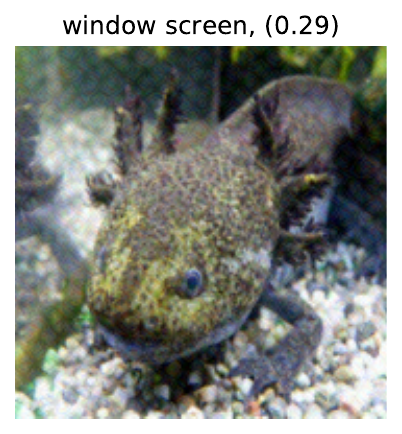}
\end{subfigure}
\begin{subfigure}{0.30\columnwidth}
\centering
\includegraphics[width=\textwidth]{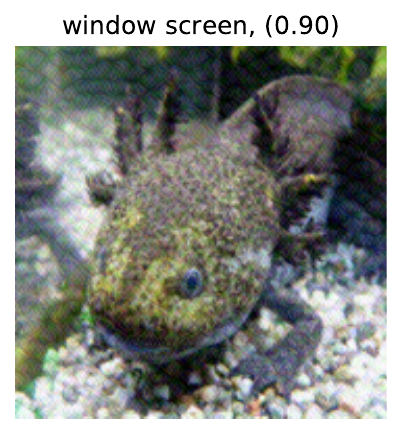}
\end{subfigure}
\end{subfigure}

\begin{subfigure}{0.98\columnwidth}
\centering
\begin{subfigure}{0.30\columnwidth}
\includegraphics[width=\textwidth]{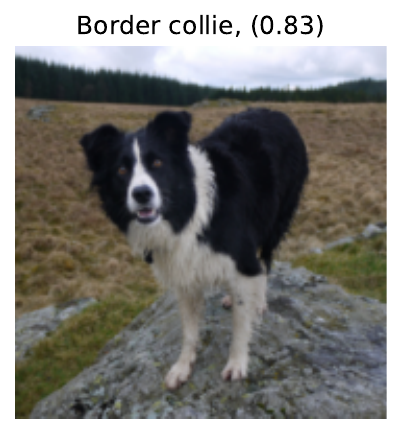}
\caption{Clear image}
\end{subfigure}
\begin{subfigure}{0.30\columnwidth}
\includegraphics[width=\textwidth]{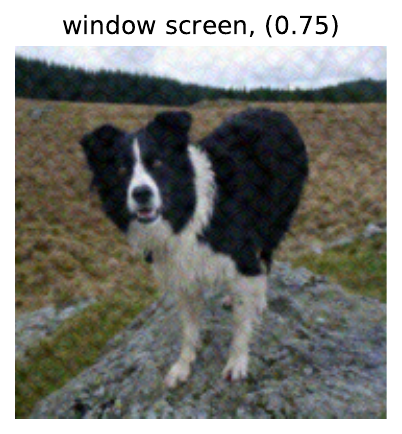}
\caption{$\epsilon=16/255$}
\end{subfigure}
\begin{subfigure}{0.30\columnwidth}
\centering
\includegraphics[width=\textwidth]{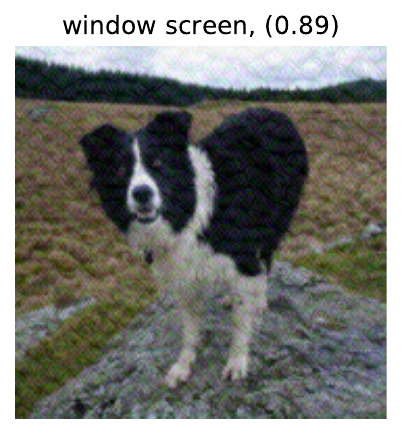}
\caption{$\epsilon=24/255$}
\end{subfigure}
\end{subfigure}

\caption{\small{Illustrations of the edge-only attacks (class = 'tench') on MobileNetv2 using different $\epsilon$. Predicted classes and confidences are shown on top. The perturbations are obtained considering the edge part with depth 1/4 of the backbone.}}
\label{f:perturbed_images}
\end{figure}

%% file: new_tables/tables_threat_model.tex
\begin{table*}[t]
\centering
\resizebox{1.\textwidth}{!}{%
\begin{tabular}{lcccc|cccc|cccc|cccc}
\toprule
 & \multicolumn{8}{c|}{\textbf{Untargeted Attack (Accuracy $\downarrow$)}} 
 & \multicolumn{8}{c}{\textbf{Targeted Attack (Success Rate $\uparrow$)}} \\
\cmidrule(lr){2-9} \cmidrule(lr){10-17}

 & \multicolumn{4}{c}{Edge CosUAP (Depth) - Sec. \ref{ss:untarget_attack}} 
 & \multicolumn{4}{c|}{\cellcolor{lightgrey}Black-Box Transfer (Attack Model)}
 & \multicolumn{4}{c}{EdgeUAP (Depth) Sec. \ref{ss:opt_uap}}
 & \multicolumn{4}{c}{\cellcolor{lightgrey}Black-Box Transfer (Attack Model)} \\

\cmidrule(lr){2-5} \cmidrule(lr){6-9} 
\cmidrule(lr){10-13} \cmidrule(lr){14-17}

Model 
& 1/4 & 2/4 & 3/4 & 4/4 
& \cellcolor{lightgrey}MNV2 & \cellcolor{lightgrey}RN50 & \cellcolor{lightgrey}VGG16 & \cellcolor{lightgrey}WR110 
& 1/4 & 2/4 & 3/4 & 4/4 
& \cellcolor{lightgrey}MNV2 & \cellcolor{lightgrey}RN50 & \cellcolor{lightgrey}VGG16 & \cellcolor{lightgrey}WR110 \\

\midrule

MNV2  
& 0.62 & 2.58 & 2.60 & 0.20 
& \cellcolor{lightred}0.70 & \cellcolor{lightgrey}11.15 & \cellcolor{lightgrey}19.93 & \cellcolor{lightgrey}14.31
& 4.26 & 4.26 & 7.04 & 67.66
& \cellcolor{lightred}97.91 & \cellcolor{lightgrey}34.39 & \cellcolor{lightgrey}32.11 & \cellcolor{lightgrey}24.90 \\

RN50  
& 0.68 & 2.58 & 1.36 & 0.86 
& \cellcolor{lightgrey}45.24 & \cellcolor{lightred}0.68 & \cellcolor{lightgrey}22.83 & \cellcolor{lightgrey}8.708
& 0.14 & 10.27 & 38.63 & 90.17
& \cellcolor{lightgrey}15.61 & \cellcolor{lightred}94.30 & \cellcolor{lightgrey}31.67 & \cellcolor{lightgrey}59.87 \\

VGG16 
& 0.82 & 1.52 & 0.82 & 0.48 
& \cellcolor{lightgrey}28.76 & \cellcolor{lightgrey}8.76 & \cellcolor{lightred}0.44 & \cellcolor{lightgrey}11.47
& 0.06 & 1.78 & 44.62 & 80.36
& \cellcolor{lightgrey}66.11 & \cellcolor{lightgrey}41.32 & \cellcolor{lightred}96.90 & \cellcolor{lightgrey}28.17 \\

WR110 
& 20.98 & 10.33 & 4.72 & 2.32 
& \cellcolor{lightgrey}63.28 & \cellcolor{lightgrey}20.04 & \cellcolor{lightgrey}34.91 & \cellcolor{lightred}1.80
& 0.16 & 7.78 & 40.94 & 74.51
& \cellcolor{lightgrey}6.08 & \cellcolor{lightgrey}36.75 & \cellcolor{lightgrey}10.89 & \cellcolor{lightred}89.22 \\

\bottomrule
\end{tabular}
}
\caption{\small{\add{Comparison of Edge-Only and Black-Box models under untargeted and targeted attacks. Classification accuracy (↓) is reported for untargeted attacks, and Targeted Success Rate (TSR) (↑) for targeted attacks, targeting class 109 (brain coral). Red cells denote the optimal transferability given by a match between surrogate and target model (white-box setting).}}}
\label{tbl:threat_model_comparisons}
\end{table*}

%% file: new_tables/table_adversarial_training.tex
\begin{table*}[t]
\centering
\setlength{\tabcolsep}{5pt}
\renewcommand{\arraystretch}{1.2}

\begin{tabular}{c|c|c|cccccccccc}
\toprule
\multirow{2}{*}{Robust Model} & \multirow{2}{*}{Clean Acc} & \multirow{2}{*}{Threat Model} 
& \multicolumn{10}{c}{Classes} \\
\cmidrule(lr){4-13}
& & & 1 & 2 & 3 & 4 & 5 & 6 & 7 & 8 & 9 & 10 \\
\midrule

\multirow{5}{*}{ResNet50} &
\multirow{5}{*}{83.28\%}
& Edge-1/4 & 54.6 & 54 & 50.8 & 50.6 & 57.6 & 56.9 & 40.1 & 64.3 & 69.1 & 57.2 \\
& & Edge-2/4 & 31.1 & 46.1 & 38.7 & 43.2 & 43.6 & 51.7 & 21.3 & 50.1 & 21.3 & 48.6 \\
& & Edge-3/4 & 24.9 & 35.8 & 35.3 & 30 & 33.6 & 36.7 & 14.7 & 34.6 & 40.1 & 44.4 \\
& & Edge-4/4 & 15.8 & 27.6 & 26.3 & 28.6 & 22.9 & 30.2 & 10 & 24.3 & 35.3 & 39.1 \\
& & \cellcolor{lightgrey}UAP        & \cellcolor{lightgrey} 18 & \cellcolor{lightgrey} 34.6 & \cellcolor{lightgrey} 25.3 & \cellcolor{lightgrey} 23.3 & \cellcolor{lightgrey} 21.7 & \cellcolor{lightgrey} 31 & \cellcolor{lightgrey} 18.8 & \cellcolor{lightgrey} 32.9 & \cellcolor{lightgrey} 32.4 & \cellcolor{lightgrey} 34.8 \\
\midrule

\multirow{5}{*}{MobileNetV2} &
\multirow{5}{*}{83.35\%}
& Edge-1/4 & 45.2 & 31 & 58.9 & 48.8 & 35.2 & 59.1 & 30.1 & 42.8 & 46 & 22.1 \\
& & Edge-2/4 & 24.8 & 22.4 & 36.8 & 35.9 & 27.3 & 37.8 & 14.8 & 29.8 & 23.2 & 19.9 \\
& & Edge-3/4 & 23.6 & 18.2 & 32.8 & 30.6 & 22 & 36.8 & 15 & 24.4 & 22.8 & 10.2 \\
& & Edge-4/4 & 18.8 & 19.6 & 31 & 30.4 & 22.2 & 31.4 & 14.1 & 21.9 & 19.4 & 7.4 \\
& & \cellcolor{lightgrey}UAP        & \cellcolor{lightgrey} 26.3 & \cellcolor{lightgrey} 20.8 & \cellcolor{lightgrey} 27 & \cellcolor{lightgrey} 32.8 & \cellcolor{lightgrey} 22.6 & \cellcolor{lightgrey} 31.7 & \cellcolor{lightgrey} 20.9 & \cellcolor{lightgrey} 27 & \cellcolor{lightgrey} 23.1 & \cellcolor{lightgrey} 14.9 \\
\midrule

\multirow{5}{*}{VGG16} &
\multirow{5}{*}{85\%} 
& Edge-1/4 & 40 & 41.9 & 26.1 & 34.8 & 36 & 43.9 & 18.2 & 37.2 & 32.2 & 33.3 \\
&  & Edge-2/4 & 25.8 & 33.2 & 17.3 & 27.4 & 19.6 & 30.7 & 9.8 & 16.8 & 16.6 & 21.7 \\
&  & Edge-3/4 & 17.1 & 23.1 & 13.8 & 24.4 & 14.1 & 26.8 & 4.2 & 15.4 & 14.9 & 10 \\
& & Edge-4/4 & 11.3 & 22.3 & 10.4 & 19.3 & 13 & 22.1 & 3.4 & 10.7 & 13.7 & 7.3 \\
& & \cellcolor{lightgrey}UAP        & \cellcolor{lightgrey} 13.6 & \cellcolor{lightgrey} 19.6 & \cellcolor{lightgrey} 12 & \cellcolor{lightgrey} 9.5 & \cellcolor{lightgrey} 16.4 & \cellcolor{lightgrey} 20 & \cellcolor{lightgrey} 12.3 & \cellcolor{lightgrey} 16 & \cellcolor{lightgrey} 16.7 & \cellcolor{lightgrey} 14.8 \\
\bottomrule
\end{tabular}
\caption{\small{\add{Evaluation of the effectiveness of adversarial training on CIFAR-10 against edge-only attacks performed with $\epsilon=16/255$. Note that we conducted targeted attacks against all 10 classes of CIFAR-10, represented in the plot with different colors, considering the various edge depths studied in previous experiments. The final bars in the plot indicate the performance of untargeted UAPs.}}}
\label{tb:adversarial_training}
\end{table*}

%% file: conclusion.tex
\section{Discussion and Conclusion}
\label{s:conc}
\add{
This work introduces a new threat model in the field of distributed learning and inference that enables edge-only adversarial attacks. In particular, we define two approaches for generating UAPs for both targeted and untargeted attacks without requiring any knowledge of the cloud-side model and the full output class distribution.

Experimental results demonstrate that our attacks can induce strong adversarial effects, even when only the initial layers of the edge model are accessible. These result in a high number of mispredictions, comparable to standard UAPs crafted with full knowledge of the model and its output class distribution. We also investigated how the crafted perturbations propagate through the cloud component, highlighting the challenge of influencing an unknown decision-making process. 

One important direction is the development of novel defense mechanisms specifically tailored to distributed inference settings, as discussed in Sec. \ref{ss:defense}. In particular, future work should address robustness at the interface between edge and cloud components. Another important direction is the extension of this attack paradigm to other domains, which is especially relevant for the IoT and sensing communities, including applications in deep signal processing.

We believe that the findings of this work contribute to a broader understanding and awareness of the threats that distributed learning systems face.
}

{\small
\section*{Acknowledgements}
This work was partially supported by project SERICS (PE00000014) under the MUR (Ministero dell'Università e della Ricerca) National Recovery and Resilience Plan funded by the European Union - NextGenerationEU.}

%% file: appendix.tex

\section{Split Point and Architectures}
\noindent In the following, we provided additional details concerning the splitting points of the tested DNNs. Subsequently, we discussed the architecture used for defining the binary classifiers $g^l$ used for the edge-only attacks.

\subsection{Selection of the split points}
\add{
Each of the considered neural networks (VGG16, ResNet50, WideResNet101, and MobileNetV2) is split under the assumption that their backbones can be schematically divided into four approximately equal depths. This hierarchical partitioning aligns with the internal structure of each model, which can be segmented into four sequential components (key layers $l_1, l_2, l_3,$ and $l_4$, as defined in Sec.~\ref{s:exp}). This setup enables the analysis of different edge depths, ranging from $1/4$ up to the full backbone, in line with related work (e.g.,~\cite{NEURIPS2023_ab003a4f_gansplit}).

To clarify the specific layers used and ensure reproducibility of the experiments, we report the exact layer paths from the torchvision models in Table~\ref{tab:layers}.
\input{new_tables/layer_table}
}

\subsection{Architecture of the classifiers}
The classifier used to distinguish between target features and other features adopts a simple architecture to prevent overfitting to a small set of targeted samples, enabling better generalization to a broader set of samples within the same target class. The pseudocode of its architecture is in Fig.~\ref{alg:pseudocode}.

For this purpose, the first layer is a convolutional module with a kernel size of 3 that takes as input the features from the selected layers. The input and output channels vary depending on the selected layers, as it corresponds to the channel size of the select edge layer $l$. The output of the convolution is then passed through an adaptive average pooling layer, allowing the model to focus on channel activation rather than spatial information. This design choice is essential to make the classifier agnostic to the position of the features (not particularly relevant in the case of universal perturbations) and focused solely on channel analysis. Finally, a fully connected layer with a hidden size of 128 is used.

The classifiers $g^l$ were trained using an Adam optimizer with default settings and a learning rate $\textit{lr} = 1 \times 10^{-3}$. The number of epochs was set to $50$, and the batch size (used for both training and attack optimization) was $100$.
Please note that, since the number of target samples in $\mathcal{D}$ was generally lower than the number of non-target samples, we used weighted sampling to balance the probability of selecting target and non-target samples during training

\add{
Please note that, regarding the computational cost of the optimization, although this is generally not the primary concern in the UAP setting (since perturbations are assumed to be computed offline and then applied at runtime), the implementation of targeted attacks requires lightweight training on edge features, whereas the formulation of CosUAP attacks does not. In both cases, the optimization process is faster than in classic white-box UAP methods because backpropagation is performed only through a part of the model’s layers.
}

\lstset{
    language=Python,
    stepnumber=1,
    numbersep=1pt,
    tabsize=2,
    showspaces=false,
    showstringspaces=false
    basicstyle=\ttfamily\small,  
    keywordstyle=\color{blue},
    commentstyle=\color{gray},
    stringstyle=\color{red},
    breaklines=true,
    columns=fullflexible,
}
\begin{figure}[ht]
\centering
\begin{subfigure}{0.98\columnwidth}
\begin{center}
\lstset{basicstyle=\ttfamily\scriptsize} %
\begin{lstlisting}
class g_Classifier(nn.Module):
    def __init__(self, n_channels, num_classes=1):
        super().__init__()
        self.conv_block = nn.Sequential(
            nn.Conv2d(n_channels, n_channels, 3, padding=1),
            nn.AdaptiveAvgPool2d(1),
            nn.Flatten()
        ) if num_channels > 1 else nn.Identity()
        self.fc_block = nn.Sequential(
            nn.Linear(n_channels, 128),
            nn.ReLU(),
            nn.Linear(128, 1)
        )
    def forward(self, x):
        return self.fc_block(self.conv_block(x))
\end{lstlisting}
\end{center}
\end{subfigure}
\caption{\small{Pseudocode in PyTorch for the binary classifier used to distinguish feature representations.}}
\label{alg:pseudocode}
\end{figure}

%% file: new_tables/layer_table.tex
\begin{table}[b]
\centering
\caption{\small{\add{Edge depths and corresponding PyTorch layers.}}}
\label{tab:layers}
\resizebox{1.\columnwidth}{!}{%
\begin{tabular}{lccccc}
\hline
\textbf{Depth} 
 & \textbf{MobileNetV2} 
 & \textbf{VGG16} 
 & \textbf{ResNet50} 
 & \textbf{WRN101} 
 & \textbf{ViT-B/S} \\
\hline
1/4 
 & \texttt{features[4]}
 & \texttt{features[9]}
 & \texttt{layer1}
 & \texttt{layer1}
 & \texttt{blocks[1]} \\

2/4 
 & \texttt{features[8]}
 & \texttt{features[16]}
 & \texttt{layer2}
 & \texttt{layer2}
 & \texttt{blocks[4]} \\

3/4 
 & \texttt{features[12]}
 & \texttt{features[23]}
 & \texttt{layer3}
 & \texttt{layer3}
 & \texttt{blocks[6]} \\

4/4 
 & \texttt{features[18]}
 & \texttt{features[30]}
 & \texttt{layer4}
 & \texttt{layer4}
 & \texttt{blocks[10]} \\
\hline
\end{tabular}
}
\end{table}